\documentclass[12pt,a4paper]{article}
\usepackage{graphicx}
\usepackage{amsmath}
\usepackage{amssymb}
\usepackage{hyperref}
\usepackage{bm}
\usepackage[height=8.8in,width=6.45in]{geometry}
\usepackage{tikz}

\numberwithin{equation}{section}

\def\tr{\mathop{\mathrm{tr}}\nolimits}
\def\diag{\mathrm{diag}}
\def\SU{\mathrm{SU}}
\def\UU{\mathrm{U}}
\def\USp{\mathrm{USp}}
\def\SO{\mathrm{SO}}
\def\PE{\mathrm{PE}}
\def\bC{\mathbb{C}}
\def\bZ{\mathbb{Z}}
\def\cN{\mathcal{N}}
\def\cP{\mathcal{P}}
\def\fg{\mathfrak{g}}

\newcommand{\ba}{\begin{array}}
\newcommand{\ea}{\end{array}}
\newcommand{\bi}{\begin{itemize}}
\newcommand{\ei}{\end{itemize}}
\def\vec#1{\bm{#1}}
\def\bea#1\eea{\allowdisplaybreaks \begin{align}#1\end{align}}
 \newcommand{\ben}{\begin{enumerate}}
\newcommand{\een}{\end{enumerate}}
\newcommand{\bean}{\begin{eqnarray*}}
\newcommand{\eean}{\end{eqnarray*}}
\newcommand{\eref}[1]{(\ref{#1})}

\newcommand{\nn}{\nonumber}

\newcommand{\BZ}{\mathbb{Z}}

\newcommand{\comment}[1]{}

\newcommand{\CI}{{\cal I}}

\newcommand{\eg}{{\it e.g.}}

\newcommand{\ud}{\mathrm{d}}

\newcommand{\tK}{\widetilde{K}}
\newcommand{\tP}{\widetilde{P}}

\newcommand{\tI}{\widetilde{{\cal I}}}

\newcommand{\be}{\begin{eqnarray}}
\newcommand{\ee}{\end{eqnarray}}
\newcommand{\bn}{\begin{enumerate}}
\newcommand{\en}{\end{enumerate}}

\def\half{\frac{1}{2}} 
\def\s{\sigma}
\def\cC{{\mathcal C}}
\def\cS{{\mathcal S}}
\def\cI{{\mathcal I}}

\def\IZ{\mathbb{Z}}

\def\a{{\vec \alpha}}
\def\b{{\vec \beta}}
\def\g{{\vec \gamma}}

\def\l{{\lambda}}
\def\m{\mu}
\def\n{\nu}
\def\G{\Gamma}
\def\L{{\vec \Lambda}}

\def\Tr{{\rm Tr}}
\def\tr{{\rm tr}}

\newcommand{\ket}[1]{|{#1}\rangle}

\begin{document}

\begin{titlepage}

\begin{flushright}
MPP-2012-153\\
UCSD-PTH-12-18 \\
IPMU-12-0219\\
UT-12-40\\
\end{flushright}

\vskip 1.5cm

\begin{center}
{\Large \bfseries
2d TQFT structure of the superconformal indices \\[.5em]
with outer-automorphism twists
}

\vskip 1.2cm

Noppadol Mekareeya$^\natural$,
 Jaewon Song$^\sharp$, 
and Yuji Tachikawa$^\flat$

\bigskip
\bigskip

\begin{tabular}{ll}
$^\natural$ &Max-Planck-Institut f\"ur Physik (Werner-Heisenberg-Institut), \\
& F\"ohringer Ring 6, 80805 M\"unchen, Deutschland \\[.5em]
$^\sharp$ &  Department of Physics, University of California, San Diego, \\
& La Jolla, CA 92093, USA\\[.5em]
$^\flat$  & Department of Physics, Faculty of Science, \\
& University of Tokyo,  Bunkyo-ku, Tokyo 133-0022, Japan, and \\
  & Institute for the Physics and Mathematics of the Universe, \\
& University of Tokyo,  Kashiwa, Chiba 277-8583, Japan
\end{tabular}

\vskip 1cm

\textbf{abstract}
\end{center}
We study the superconformal indices of 4d theories coming from 6d $\cN=(2,0)$ theory of type $\Gamma$ on a Riemann surface, with the action of the outer-automorphism $\sigma$ in the trace. We find that the indices are given by the partition function of a deformed 2d Yang-Mills on the Riemann surface with gauge group $G$ which is S-dual to the subgroup of $\Gamma$ fixed by $\sigma$.
In the 2-parameter deformed version, we find that it is governed \emph{not} by Macdonald polynomials of type $G$, but by Macdonald polynomials associated to twisted affine root systems. 

\medskip
\noindent

\bigskip
\vfill
\end{titlepage}

\setcounter{tocdepth}{2}
\tableofcontents

\section{Introduction}
Consideration of the 6d $\cN=(2,0)$ theory of type $A$ on the spacetime of the form $S^3\times S^1\times \cC_2$ recently led the authors of \cite{Gadde:2009kb,Gadde:2011ik,Gadde:2011uv,Gaiotto:2012xa} to the following observation. The compactification of the 6d theory on $\cC_2$ leads to a 4d $\cN=2$ theory often called a theory of class $\cS$ \cite{Gaiotto:2009we, Gaiotto:2009hg}.  In the low energy limit the theory becomes superconformal, and then its partition function on $S^3\times S^1$ is called the superconformal index \cite{Kinney:2005ej, Romelsberger:2005eg}. This can be explicitly calculated when the theory of class $\cS$ in question has a Lagrangian description, and equals the partition function of deformed 2d Yang-Mills with gauge group $\SU(n)$ on $\cC_2$. In the setup, we have three parameters $(p,q,t)$\footnote{Here and in the following, we use the notation of \cite{Gaiotto:2012xa} for $(p,q,t)$.}. In the one-parameter subspace $(p=0,q,t=q)$, the deformation is the standard $q$-deformation, and in the two-parameter subspace $(p=0,q,t)$, the deformation is described by Macdonald polynomials of type $A$.
The full three-dimensional parameter space corresponds to an elliptic generalization to Macdonald polynomials, about which not much is known, except some developments reported in \cite{Gaiotto:2012xa}. 

The same setup was analyzed in \cite{Kawano:2012up,Fukuda:2012jr} by exchanging the order of the compactification. Let us first compactify the 6d theory on $S^1$. We then have 5d maximally-supersymmetric Yang-Mills theory on $S^3\times \cC_2$. We can perform supersymmetric localization of this system to show that it reduces to the 2d deformed Yang-Mills theory on $\cC_2$.  Currently it is only possible to analyze the one-parameter subspace  $(p=0,q,t=q)$ in this approach, where the $q$-deformation arises from the one-loop determinants of the Kaluza-Klein modes along $S^3$. 

\begin{table}[ht]
\[
\begin{array}{r||cccc|c}
\Gamma & A_{2\ell-1} &  D_{\ell+1} & D_4 &  E_6  & A_{2\ell} \\
\hline
d & 2 & 2 &  3 & 2 & 2 \\
\hline
G & B_\ell & C_\ell & G_2 & F_4 & C_\ell \\
\hline
\text{Macdonald}& C_\ell^\vee & B_\ell^\vee & G_2^\vee & F_4^\vee   & (C_\ell^\vee,C_\ell)
\end{array}
\]
\caption{The type of the 6d theory, the order of the outer-automorphism twists, the resulting 5d gauge group, and the type of the Macdonald polynomials governing the deformation.\label{twist-table}}
\end{table}

 A 6d $\cN=(2,0)$ theory is labeled by a simply-laced Dynkin diagram $\Gamma$, and the graph automorphism group of $\Gamma$ is a discrete symmetry group of the theory. We call this symmetry group the outer-automorphism group. One of the mysterious features of the 6d theory is that when we compactify it on $S^1$ with an outer-automorphism twist $\sigma$, the resulting 5d maximally-supersymmetric Yang-Mills has the gauge group $G$ which is \emph{S-dual to} the subgroup invariant under $\sigma$ of the simply-laced group of type $\Gamma$, see Table~\ref{twist-table}. This feature was derived using various nonperturbative dualities \cite{Vafa:1997mh,Tachikawa:2011ch}, since there is still no universally accepted Lagrangian description of the 6d theory.
 
In this article, we extend the analysis of the superconformal index of the theory of class $\cS$ of type $A$, in the one- and two-parameter subspaces, to the case when we have an outer-automorphism twist $\sigma$ around $S^1$.\footnote{For $\cC_2=T^2$, it was already considered in \cite{Zwiebel:2011wa}. Note also that the outer-automorphism twist we employ in this paper is around $S^1$, not around cycles on $\cC_2$ as was in \cite{Tachikawa:2009rb,Tachikawa:2010vg}.} The class $\cS$ theory has sometimes a Lagrangian description, and the superconformal index can be calculated in the weakly-coupled regime. Therefore, our analysis can shed light on the mechanism how the group $G$ S-dual to the $\sigma$-invariant subgroup arises.  As we will see, the twist $\sigma$ does not restrict the modes to the subspace invariant under $\sigma$. Rather, it enforces the modes exchanged by $\sigma$ to appear always in pairs, effectively producing a particle whose charge is described by the group $G$ S-dual to the $\sigma$-invariant subgroup. Thus we will find a deformed 2d Yang-Mills theory with gauge group $G$ on $\cC_2$.

We will also find that the two-parameter version is not just obtained by considering Macdonald polynomials of type $G$.
The Macdonald polynomials are parameterized by (possibly-non-reduced) affine root systems, which are \emph{not} in one-to-one correspondence with non-affine Dynkin diagrams parameterizing $G$.  The type of Macdonald polynomials which appear in the description of the 2d theory is summarized in Table~\ref{twist-table}. In particular, for $\Gamma=A_{2\ell}$ and $\bZ_2$ twist, we need the polynomials associated to $(C_\ell^\vee,C_\ell)$, which are called the Koornwinder polynomials.

Any Riemann surface $\cC_2$ is constructed from cylinders and three-punctured spheres.
The superconformal index without the twist, in the two parameter case, is known to have the structure \begin{equation}
\cI=\sum_{\vec\lambda} \frac{\prod_I K_{\Lambda_I}(\vec{a}_I) \underline{P}_{\vec\lambda}(\vec{a}_I t^{\Lambda_I})}{ K_{\vec\rho} \underline{P}_{\vec\lambda}(t^{\vec\rho})} .\label{generalindex}
\end{equation}  Here, $\vec\lambda$ runs over the irreducible representation of the group $\Gamma$,
$\underline{P}_{\vec\lambda}$ is the normalized Macdonald polynomial of the corresponding type, 
$\vec{\Lambda}_I$ specifies the type of the $I$-th puncture,
$\vec{\rho}$ is the type of the null puncture, 
$\vec a_I$ is the fugacity of the flavor symmetry of the $I$-th puncture, 
$\vec  a t^{\Lambda}$ is determined by the type of the puncture $\vec\Lambda$, and 
$K_{\vec\Lambda}(\vec a)$ is a $\lambda$-independent prefactor given by an infinite product \begin{equation}
K_{\vec\Lambda}(\vec a)=\prod_{n=0}^\infty\prod_{s} \frac{1}{1-z_{\Lambda,s}(\vec a,q,t)q^n},\label{above}
\end{equation} where $z_s(\vec a,q,t)$ is a monomial of the indicated arguments.

In the twisted case, we analyze cylinders for all types of 6d theory, and three-punctured spheres of type $A$.  Assuming that the twist is of order $2$, we will find that the index is given by \begin{equation}
\tilde\cI=\sum_{\vec\lambda} \frac{\prod_I \tilde K_{\Lambda_I}(\tilde{\vec{a}}_I) \underline{\tilde{P}}_{\vec\lambda}(\tilde{\vec{a}}_I t^{\Lambda_I})}{ \tilde K_{\vec\rho} \underline{\tilde P}_{\vec\lambda}(t^{\vec\rho})} . \label{twistedgeneralindex}
\end{equation} Here, $\vec\lambda$ now runs over the irreducible representation of the group $G$,
$\underline{\tilde P}_\lambda$ is the normalized Macdonald polynomial of the type listed in Table~\ref{twist-table},
$\tilde{\vec a}_I$ is the fugacity of the flavor symmetry of the $I$-th puncture modified so that it is compatible with the twist,
and $\tilde K_{\vec\Lambda}(\tilde{\vec a})$ is now given by an infinite product \begin{equation}
\tK_{\vec\Lambda}(\tilde{\vec a})=\prod_{n=0}^\infty\prod_{s} \frac{1}{1\pm z_{\Lambda,s}(\tilde{\vec a},q,t)q^n},
\end{equation} where $z_{\Lambda, s}(\vec a,q,t)$ is the \emph{same} monomial of the indicated arguments as in \eqref{above}. 
The choice of a $+$ or a $-$ sign depends on the monomial $z_{\Lambda, s}$; the prescription is given in Sec. \ref{sec:withtwists}.
We will provide ample pieces of evidence supporting the validity of this formula.

The formula is rather complicated but the physical interpretation is straightforward: the outer-automorphism twist $\sigma$ acts in a natural way on each factor of \eqref{generalindex}. Namely, it modifies the type of the orthogonal polynomial, and it just acts by $\pm 1$ on the Fock space generators contributing to \eqref{above}. This strongly suggests that there is a physical frame where each factor in \eqref{generalindex} and \eqref{twistedgeneralindex} has a simple physical interpretation. 

The rest of the paper is organized as follows.  In Sec.~\ref{TQFT}, we review the definition of the superconformal indices of 4d theories, and its TQFT interpretation when the 4d theory considered is of class $\cS$.  In Sec.~\ref{vector}, we study the contribution from the vector multiplets, and show how the gauge group $G$ S-dual to the $\sigma$-fixed subgroup of $\Gamma$ appears from a perturbative calculation. In Sec.~\ref{hyper}, we study the twisted superconformal indices of free hypermultiplets coming from three-punctured spheres, and check the validity of the formula \eqref{twistedgeneralindex}. We also perform the check of the agreement of the superconformal index on both sides of the Argyres-Seiberg duality. We have a few appendices detailing the niceties. 

\section{Twisted superconformal indices and TQFT}\label{TQFT}
In this section, we review the superconformal indices and their 2d TQFT structure and then define the twisted superconformal index.  

\subsection{Superconformal indices and topological field theory}

The superconformal index of $\cN=2$ SCFT is defined as \cite{Gadde:2011uv, Gaiotto:2012xa}
\be
 \cI &=& \Tr (-1)^F p^{\half (E+2j_1 - 2R - r)} q^{\half (E-2j_1 - 2R - r)} t^{R+r} \prod_i x_i^{f_i}  \nn \\
  &=& \Tr (-1)^F \left( \frac{t}{pq} \right)^r p^{j_2 + j_1} q^{j_2 - j_1} t^R \prod_i x_i^{f_i} , \label{n2index}
\ee
where $E$ is the conformal dimension, $(j_1, j_2)$ are the Cartans of the Lorentz rotation $\SU(2)_1 \times \SU(2)_2$, $r$ is the $\UU(1)_R$ generator and $R$ is the Cartan generator of $\SU(2)_R$. The $f_i$ are the Cartan generators of the flavor symmetry group. 
The trace is over the states with $\Delta \equiv E - 2j_2 - 2R + r = 0$ upon radial quantization. 

The single letter indices for a vector multiplet and a half-hypermultiplet are given by
\be
 f^\text{vect} &=& -\frac{p}{1-p} - \frac{q}{1-q} + \frac{pq/t - t}{(1-q)(1-p)} , \\
 f^{\half\text{hyp}} &=& \frac{\sqrt{t} - pq/\sqrt{t}}{(1-q)(1-p)} . \label{single}
\ee
For an $\cN=2$ theory with Lagrangian description, one can obtain the superconformal index first by writing a single letter index and then put it in a plethystic exponential
\be
 \cI = \PE \left[\sum_i f_{R_i} (p, q, t) \chi_{R_i} (U) \right]_{p, q, t, U}  \label{PE} , 
\ee
where $f_{R_i}$ are the single letter indices for a $\cN=2$ multiplet in representation $R_i$ of the gauge group. 
Here, the symbol $\PE$ stands for the plethystic exponential, defined by 
\begin{equation}
\PE \left[\sum_{n\geq 1} a_n t^n \right]_{t} = \prod_{n\ge 1}(1-t^n)^{-a_n}.
\end{equation} We often omit the subscripts of $\PE$ signifying the variable(s) with respect to which the exponential is taken.
Subsequently, we also often use the following expression: 
\bea
\PE \left[u \sum_{n\ge 1} a_n t^n \right]_{u=-1} = \prod_{n\ge 1}(1+t^n)^{-a_n};
\eea
this is also known in the literature as the fermionic plethystic exponential (see \eg, \cite{Feng:2007ur}).

The index $\cI$ becomes the 2-parameter Macdonald index if we take $p = 0$, and becomes the Schur index when we take $t=q$ as well. Our focus on this paper is the Macdonald index with 2-parameters
\be
 \cI (p, q) = \Tr (-1)^F q^{E-2R - r} t^{R+r} = \Tr(-1)^F q^{-2j_1} t^{R+r} , 
\ee
where the trace is over the $\frac{1}{4}$-BPS states with $\Delta' \equiv E + 2j_1- 2R - r = 0$ in addition to $\Delta = 0$. 

It is straight-forward to evaluate the superconformal index when the Lagrangian description of the theory is available. For example, indices for all the theories in class $\cS$ of type $A_1$ can be evaluated. It is shown in \cite{Gadde:2009kb} that the indices can be thought of as correlation functions of a 2-dimensional topological field theory on $\cC_2$ used to construct the theory. Moreover, the TQFT structure enables us to evaluate the indices for non-Lagrangian theories as well  \cite{Gadde:2011ik, Gadde:2011uv}. The idea of TQFT as applied to this setup is as follows:  
\bn
 \item Any TQFT correlators can be obtained from knowing the three point functions and propagators. A 3-point function can be written as
 \be
  I_{{\vec\Lambda}_1, {\vec\Lambda}_2, {\vec\Lambda}_3} (\vec{a}, \vec{b}, \vec{c}) = \sum_{\lambda \m \n} C^{\lambda \m \n} f_{\lambda}(\vec a; \L_1) f_{\m}(\vec b; \L_2) f_{\n}(\vec c; \L_3) , 
 \ee
where $C^{\l \m\n}$ is called the structure constant. The propagator is given by 
\be
\eta(\vec{a}, \vec{b}) = I^V (\vec{a}) \delta(\vec{a}, \vec{b}^{-1}) , 
\ee
where $I^V(\vec a)$ is given by the index of a vector multiplet and $\delta$ is a Dirac delta function. One can glue the three point functions with propagators to form arbitrary correlators. 

 \item Choose the basis of functions $\{ f^{\m} \}$ to be orthogonal under the measure involving the propagator which is the index of a vector multiplet:
 \be
 \eta_{\m \n} = \int [d\vec{a}][d\vec{b}] \eta(\vec{a}, \vec{b}) f_\m (\vec{a}) f_\n (\vec{b}) 
 = \int [d\vec{a}] I^V(\vec{a}) f_{\m}(\vec{a}) f_{\n }(\vec{a}^{-1}) = \delta_{\m \n} ,
 \ee
where $[d\vec{a}]$ is the Haar measure of the gauge group. Once this is done, we can freely move the indices of the structure constant upward or downward. 

 \item A nontrivial fact is that the structure constant $C^{\l\m\n}$ can be made diagonal, which means that only $C^{\m\m\m}$ are non-zero. This assures the associativity of TQFT, which in turn guarantees S-dual invariance of the index. Also, this justifies our notation of the structure constant which is independent of the type of punctures on $\cC_2$. 

 \item Write $f_{\l}(\vec{a}; {\vec \Lambda}) = K_{{\vec \Lambda}} (\vec{a}) P_{\l} (\vec{a} t^{{\vec \Lambda}} )$ where $P_{\l}(\vec a)$ is some known symmetric polynomial orthogonal under certain measure $\Delta'(\vec a)$ and $\vec a t^{{\vec \Lambda}}$ is specified by the type of puncture ${\vec \Lambda}$. 

 \item There is a type of puncture, $\vec\rho$, which corresponds to the absence of any puncture. Then the consistency requires that $C^\mu=(K_{\vec\rho} P_{\vec\mu}(t^{\vec\rho}) )^{-1}$

\en
Here we omitted the dependence on the fugacities $(p, q, t)$ from the notation. 

Since we know how the orthogonal polynomial $P_{\l} (\vec{a})$ generalizes to higher rank, we can write down a conjecture for the superconformal indices of the theories in class $\cS$.
The non-trivial piece of information we need to determine is then  the prescription for $K_{\vec \Lambda} (\vec{a})$ and $\vec{a} t^{\vec \Lambda}$ for generic type of punctures. 

Once they are given, we can write the superconformal index of a theory in class $\cS$ defined by a Riemann surface $\cC_2$ of genus $g$ and $s$ punctures of type ${\vec\Lambda}_{1, \ldots, s}$ as 
\be
 \cI = \sum_{\vec\l} \frac{ \prod_{I=1}^s K_{\L_I} (\vec{a}_I) P_{\l} (\vec{a}_I t^{\L_I})}{(K_{\vec\rho} P_{\vec\lambda} (t^{\vec\rho})) ^{2g-2+s}}. 
\ee

In the following, we will define the superconformal index with outer-automorphsim twist and then write it as a correlation function of a 2d topological field theory as the untwisted case. 

\subsection{Superconformal indices with outer-automorphism twist}

We define the twisted superconformal index as 
\be
 I = \Tr(-1)^F \cP_\sigma q^{m} t^{n} \prod_{i} x_i^{f_i} . 
\ee
where $\cP_\sigma$ acts as outer-automorphism $\s$ on the gauge group $\Gamma$ and $m = -2j_1, n = R+r$. Upon twisting, the index should get contributions only from the eigenstates of $\cP_\sigma$. 

Let $\Gamma$ be the gauge group of the theory, and $G$ be the group obtained by folding the Dynkin diagram of $\Gamma$ or the S-dual of $\s$-invariant subgroup. We assume $\Gamma$ to be simply-laced. Let's denote the number of simple roots of $\Gamma$ invariant under $\s$ to be $r'$ and the rank of $\Gamma$ to be $r$. Write $r = r' + dr''$, where $d$ is the order of the outer-automorphism group of $\Gamma$. 
Then the rank of $G$ is given by $r' +  r''$  For $\Gamma = A_\ell, D_\ell, E_6$ the outer-automorphism group is given by $\IZ_2$ and for $D_4$, one can have $\IZ_3$.   

Suppose we label non-gauge invariant states contributing to the Macdonald index as $v_{\vec\alpha; m, n}$ where $m, n$ are the eigenvalues of the corresponding operators, and $\vec\alpha$ is a weight of the gauge group. 
At the end of day, we should pick up gauge invariant states only. We will discuss it later in this section. For the time being, we regard the gauge symmetry as a flavor symmetry and introduce the fugacities $x$ associated to it to the index. We will use the multiplicative notation for the fugacities. So, for a vector ${\alpha}_i$ and fugacities $x_i$,  $x^{\vec \alpha}$ stands for the product $\prod_i x_i{}^{{\alpha}_i}$.

The outer-automorphism $\sigma$ becomes an operator $\cP_\sigma$ acting on the Fock space. 
Here we consider its actions on the modes which transform as adjoints, i.e.~$\vec\alpha\in \Delta$, where $\Delta$ is the set of roots. Other representations can be considered similarly, and we will encounter them in later sections.
If the root ${\vec \alpha}$ is sent to ${\vec \beta}$ by $\sigma$, the mode $v_{{\vec \alpha};m,n}$ is sent to $v_{{\vec \beta};m,n}$. The effect of $\cP_\sigma$ to the index for the case of $\G \neq A_{2\ell}$ and $\G = A_{2\ell}$ should be treated separately. 
This distinction arises because when $\vec\beta = \sigma(\vec\alpha)\neq \vec\alpha$, $\vec\beta$ and $\vec\alpha$ are orthogonal, \emph{unless} $\G=A_{2\ell}$.

\paragraph{For $\Gamma\neq A_{2\ell}$:} 
Let  $\Gamma$ be one of $A_{2\ell-1}$, $D_{n+1}$, $D_4$ or $E_6$.  
Then the trace with the insertion of $\cP_\sigma$ is given by the product of the following factors: (Note that $x$ is now restricted to be invariant under $\sigma$.)\begin{itemize}
\item If ${\vec \alpha}$ is invariant under $\sigma$, we just have the same contribution $x^\a q^m t^n$.
\item If $\sigma$ is $\bZ_2$ and exchanges ${\vec \alpha}$ and ${\vec \beta}$, we need to study the trace of $\cP_\sigma$ acting on the states \begin{equation}
(v_{{\vec \alpha};m,n})^k (v_{{\vec \beta};m,n})^l .
\end{equation} The trace only receives the contribution from the states with $k=l$, so it's a sum over \begin{equation}
(v_{{\vec \alpha};m,n}v_{{\vec \beta};m,n})^n . 
\end{equation} 
For each such state we get the contribution $x^{\a+\b} q^{2m} t^{2n}$.

\item If $\sigma$ is $\bZ_3$ and permutes ${\vec \alpha}\to {\vec \beta}\to{\vec \gamma}\to {\vec \alpha}$, similarly it gives $x^{\a+\b+\g} q^{3m} t^{3m}$. 
\end{itemize}

\begin{figure}
\[
\vcenter{\hbox{
\begin{tikzpicture}[x=15pt,y=15pt]
\filldraw (0,0) rectangle +(1.1,1.1);
\draw (1.5,0) rectangle +(1.1,1.1);
\filldraw (3,0) rectangle +(1.1,1.1);

\draw (0,1.5) rectangle +(1.1,1.1);
\filldraw (1.5,1.5) rectangle +(1.1,1.1);
\draw (3,1.5) rectangle +(1.1,1.1);

\filldraw (0,3) rectangle +(1.1,1.1);
\draw (1.5,3) rectangle +(1.1,1.1);
\filldraw (3,3) rectangle +(1.1,1.1);

\draw[thick] (5,5) -- (-1,-1);
\draw[stealth-stealth,ultra thick] (5,4) -- (4,5);

\end{tikzpicture}
}}
\qquad\qquad
\vcenter{\hbox{
\begin{tikzpicture}[x=15pt,y=15pt]
\draw (0,0) rectangle +(1.1,1.1);
\filldraw (1.5,0) rectangle +(1.1,1.1);
\draw (3,0) rectangle +(1.1,1.1);
\filldraw (4.5,0) rectangle +(1.1,1.1);

\filldraw (0,1.5) rectangle +(1.1,1.1);
\draw (1.5,1.5) rectangle +(1.1,1.1);
\filldraw (3,1.5) rectangle +(1.1,1.1);
\draw (4.5,1.5) rectangle +(1.1,1.1);

\draw (0,3) rectangle +(1.1,1.1);
\filldraw (1.5,3) rectangle +(1.1,1.1);
\draw (3,3) rectangle +(1.1,1.1);
\filldraw (4.5,3) rectangle +(1.1,1.1);

\filldraw (0,4.5) rectangle +(1.1,1.1);
\draw (1.5,4.5) rectangle +(1.1,1.1);
\filldraw (3,4.5) rectangle +(1.1,1.1);
\draw (4.5,4.5) rectangle +(1.1,1.1);

\draw[thick] (6.5,6.5) -- (-1,-1);
\draw[stealth-stealth,ultra thick] (6.5,5.5) -- (5.5,6.5);

\end{tikzpicture}

}}
\]
\caption{The $\bZ_2$ action \eqref{explicitZ2actionAodd} and \eqref{explicitZ2actionAeven}  on the matrix. The sign is plus or minus if the entry is white or black, respectively. \label{action}}
\end{figure}
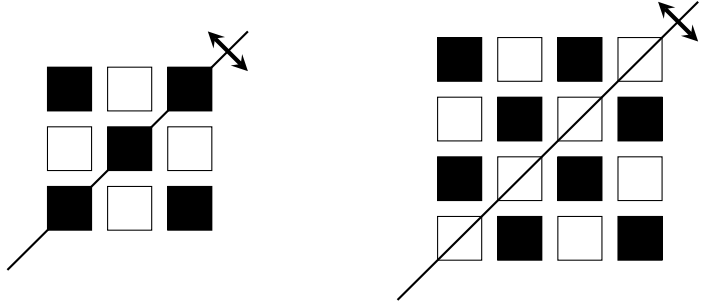

Here we record the explicit action of $\sigma$ on $A_{2\ell-1}$, as we will need it later.
Let us call the entries of $2\ell$-by-$2\ell$ matrix $e_{ij}$, where $i,j=-\ell,-(\ell-1), \ldots, -1,  1, 2, \ldots, \ell-1, \ell$. Then we pick $\sigma$ to act  by \begin{equation}
e_{i,j} \to - (-1)^{s(i)-s(j)} e_{-j,-i}\label{explicitZ2actionAodd}
\end{equation} where \begin{equation}
s(i)=\begin{cases}
i, & (i>0) \\
i+1. & (i<0)
\end{cases}
\end{equation}
 The invariant subalgebra is $\USp(2\ell)$, i.e~$C_\ell$. Its S-dual is $B_\ell$.
Note that the action \eqref{explicitZ2actionAodd} flips the matrix along its anti-diagonal, and the sign is given by the checker board pattern so that it is $-1$ on the diagonal, see Fig.~\ref{action}.

\paragraph{For $\Gamma=A_{2\ell}$ :} this case needs to be treated separately. Let us call the entries of $(2\ell+1)$-by-$(2\ell+1)$ matrix $e_{ij}$, where $i,j=-\ell,-(\ell-1), \ldots, -1, 0,  1, 2, \ldots, \ell-1, \ell$. Then one particularly nice outer automorphism $\sigma$ is given by \begin{equation}
e_{i,j} \to - (-1)^{i-j} e_{-j,-i}.\label{explicitZ2actionAeven}
\end{equation} The invariant subalgebra is $\SO(2\ell+1)$, i.e.~$B_\ell$. Its S-dual is $C_\ell$.
See Fig.~\ref{action}.
We let the Cartan subalgebra to be formed by $e_{i,i}$. Then the invariant part is given by \begin{equation}
a_i = e_{i,i}=-e_{-i,-i}
\end{equation} for $i=1,\ldots, n$. $e_{0,0}$ is forced to be zero.

Let's denote by $v_{i,j}$ the modes corresponding to $e_{i,j}$, where $i\neq j$. We suppress the dependence on $m, n$ for simplicity. 
The mode $v_{i,j}$ has charge $a_i-a_j$, where we define $a_{-i}=-a_i$.
Then $\cP_\sigma$ acts on $v_{i,j}$ by \begin{equation}
v_{i,-i} \to - v_{i,-i} 
\end{equation} for $i=-n,\ldots, n$ (excluding $i=0$) and by \begin{equation}
v_{i,j} \to -v_{-j,-i} 
\end{equation} for $i\ne -j$.   Then the combinations of $v$'s contributing to the trace are 
\begin{equation}\label{A2nCharges}
\begin{array}{c|ccc}
& \text{charge} & \text{action of }\cP_\sigma & \text{index} \\
\hline
v_{i,-i} & 2a_i & - & -x_i^2 q^m t^n  \\
v_{i,0} v_{0,-i} & 2 a_i& + & x_i^2 q^{2m} t^{2n} \\
\hline
v_{-i,i} & -2a_i & - & -x_i^{-2} q^m t^n \\
v_{-i,0} v_{0,i} & -2a_i& + & x_i^{-2} q^{2m} t^{2n} \\
\hline
v_{i,j} v_{-j,-i} & 2(a_i - a_j) & + & x_i^2 x_j^{-2} q^{2m} t^{2n}\\
v_{i,-j} v_{j,-i} & 2(a_i + a_j) & + & x_i^2 x_j^2 q^{2m} t^{2n} \\
v_{-i,j} v_{-j,i} & 2(-a_i - a_j) & + & x_i^{-2} x_j^{-2} q^{2m} t^{2n}
\end{array}
\end{equation} where $i$ and $j$ are both positive. 

\subsubsection{Haar measures} \label{haar}
In order to pick up the gauge invariant states, we need to integrate over the index obtained from non-gauge invariant states with the appropriate Haar measure. Let us recall how the Haar measure arises in the calculation of the superconformal index: 
\begin{equation}
\int [\ud {\vec x}]_\Gamma  =\frac{1}{\# W}\int \prod_{a=1}^r \frac{dx_a}{2\pi i x_a}  \prod_{{\vec \alpha} \in \Delta_\G} (1-x^{\vec \alpha})  \label{Haar}
\end{equation}  where the product is over the  roots ${\vec \alpha}$ of $\Gamma$ and $\#W$ is the order of the Weyl group of type $\Gamma$ and the integration is over the unit circle in a complex plane. 
The point is that the Wilson line $x_a$  around $S^1$ of the $\Gamma$ gauge field, gauge-fixed to lie in the Cartan subgroup, is a zero mode to be integrated over. Then the off-diagonal elements of vector multiplets of $\Gamma$, independent along $S^3$, give one-loop fluctuations, by creating a Fock space of Fadeev-Popov ghosts. 
Say a root ${\vec \alpha}$ gives a fermionic raising operator $b_{\vec \alpha}$. Then we have a pair of states 
\begin{equation}
\ket{\psi}, \quad
b_{\vec \alpha} \ket{\psi}, 
\end{equation} 
for any state $\ket{\psi}$. 
The Wilson line works as a chemical potential, and a mode $b_{\vec \alpha}$ contributes to the partition function by $x^{\vec \alpha}$.
Then the contribution from $v_{\vec \alpha}$ to the partition function is $(1-x^{\vec \alpha})$, accounting for a factor in \eqref{Haar}.

Let us now study how the Haar measure of the $\Gamma$ gauge group changes if we include the twist by $\sigma$. We will discuss the case of $\G \neq A_{2\ell}$ and $\G = A_{2\ell}$ separately as we have discussed. 

\paragraph{For $\Gamma\neq A_{2\ell}$:} 
Now, the Haar measure \eqref{Haar} becomes 
\begin{equation}
\int d\vec{x} \prod_{{\vec \alpha}=\sigma({\vec \alpha})} (1-x^{{\vec \alpha}}) \prod_{{\vec \alpha}\leftrightarrow{\vec \beta}}(1 - x^{{\vec \alpha}+{\vec \beta}} )
\end{equation} when $\sigma$ is order 2, or 
\begin{equation}
\int d\vec{x} \prod_{{\vec \alpha}=\sigma({\vec \alpha})}(1- x^{{\vec \alpha}})  \prod_{{\vec \alpha}\to{\vec \beta}\to{\vec \gamma}} (1 - x^{{\vec \alpha}+{\vec \beta}+{\vec \gamma}} )
\end{equation} when $\sigma$ is order 3. In both cases, the integral is over the $\sigma$-invariant subspace of the Cartan subgroup of $\Gamma$.
As is well known and is explained in the Appendix A of  \cite{Keller:2011ek}, the expressions  above become \begin{equation}
\int [d \vec{x}]_G = \int d\vec{x}\prod_{{\vec \alpha} \in \Delta_G} (1-x^{\vec \alpha})
\label{foo}
\end{equation} where ${\vec \alpha}$ runs over the roots of $G$, and the integral of $x$ is over the Cartan of $G$.
Thus we see that, in some sense, the outer-automorphism $\sigma$ binds the mode $v_{\vec \alpha}$ and $v_{\vec \beta}$ to a different mode $v_{\vec \alpha} v_{\vec \beta}$, to produce the charge of the gauge group $G$ which is S-dual to the subgroup invariant under $\sigma$. This appearance of  the Haar measure for $G$ can be studied from another point of view, see Appendix~\ref{propagator}.

\paragraph{For $\Gamma=A_{2\ell}$ :}
The first two modes of \eqref{A2nCharges} give the contribution \begin{equation}
(1-x_i^{2})(1+x_i^{2})=(1-x_i^{4}),
\end{equation} the second two modes \begin{equation}
(1-x_i^{-2})(1+x_i^{-2})=(1-x_i^{-4}),
\end{equation} and the last three modes give \begin{equation}
\prod_{i > j,\pm,\pm} (1-x_i^{\pm 2} x_j^{\pm 2}).
\end{equation} The total contribution is then \begin{equation}
\prod_{{\vec \alpha}} (1-(x^2)^{\vec {\vec \alpha}})\label{bar}
\end{equation} where \begin{equation}
{\vec \alpha} = \pm 2a_i, \quad \pm a_i \pm a_j
\end{equation} 
which is the measure factor for $C_\ell$, as it should be. 
Note that the contribution for the long root $2a_i$ of $C_\ell$ comes from the combination of the modes $e_{i,-i}$ and $e_{i,0} e_{0,-i}$.  Note also that the fugacity $x^2$ is rescaled by a factor of two, whereas when the $C_\ell$ Haar measure arises from $\Gamma=D_{n+1}$, there is no such rescaling, see \eqref{foo}.

\section{Contributions from vector multiplets} \label{vector}
\subsection{Without  outer-automorphism twist}
Let us first study the contribution from the vector multiplets, or equivalently the cylinder in the Riemann surface. 
A vector multiplet of the gauge group $G$  gives \begin{equation}
\int [\ud {\vec x}]_G \PE \left[\frac{-p-q+2pq + pq/t -t }{(1-p)(1-q)} \chi_\text{adj}(\vec x)\right]_{p,q,t,x} \label{measure3}
\end{equation} as can be seen from \eqref{single} and \eqref{PE}.
Here, $[\ud {\vec x}]_G$ is the Haar measure for the group $G$ as in the last section.

The two-parameter version of the measure is obtained by setting $p=0$: \begin{equation}
\int [\ud {\vec x}] \PE \left[\frac{-q -t }{1-q} \chi_\text{adj}(\vec x) \right] \label{measure2}
\end{equation} and the one-parameter version is obtained by further setting $q=t$: \begin{equation}
\int [\ud {\vec x}] \PE \left[\frac{-2q }{1-q} \chi_\text{adj}(\vec x) \right]. \label{measure1}
\end{equation} 

In the superconformal index, it is important to obtain a orthogonal measure under \eqref{measure3}, \eqref{measure2}, or \eqref{measure1}. But note that \eqref{measure1} is just \begin{equation}
\int [\ud {\vec x}] \PE \left[\frac{-q }{1-q} \chi_\text{adj}(\vec x) \right]^2.
\end{equation} Therefore, the orthogonal polynomials under it are just \begin{equation}
\PE \left[\frac{+q }{1-q} \chi_\text{adj}(\vec x) \right] \chi_{\vec \lambda}(\vec x),
\end{equation} where $\chi_{\vec \lambda}(\vec x)$ is the character of an irreducible representation (irrep) of highest weight ${\vec \lambda}$.

For the two-parameter case \eqref{measure2}, a way out is to use  Macdonald polynomials $P_{\vec \lambda}$ of type $G$,  which are defined as the orthogonal polynomials  under the measure \begin{equation}
\int [\ud {\vec x}] \prod_{n=0}^\infty \prod_{{\vec \alpha}}\frac{1-q^{n+1} x^{\vec \alpha}}{1-t q^n x^{\vec \alpha}}.\label{simplylacedmacdonaldmeasure}
\end{equation} 
Now rewrite \eqref{measure2} as (where $r$ is the rank of $G$) \begin{equation}
\eqref{measure2}=\PE \left[\frac{-q+t}{1-q}\right]^{r} \int [\ud {\vec x}] \PE \left[\frac{-q+t}{1-q}\sum_{{\vec \alpha}\in\Delta} x^{\vec \alpha}\right] \PE \left[\frac{-t}{1-q}\chi_\text{adj}(\vec x) \right]^2.\label{ooo}
\end{equation}
This makes it clear that dressed Macdonald polynomials \begin{equation}
K_\text{full}(\vec x)P_{\vec \lambda}(\vec x)
\end{equation} where \begin{equation}
K_\text{full}(\vec x)=\PE \left[\frac{t}{1-q}\chi_\text{adj}(\vec x) \right]
\end{equation} 
 are orthogonal under the two-parameter vector-multiplet measure \eqref{measure2}. 
The factor $K_\text{full}(\vec x)$ will recur as the prefactor for the full puncture later.

\subsection{With outer-automorphism twist} \label{vectortwist}

\subsubsection{Macdonald measures}
The analysis of the whole vector multiplet measure goes almost verbatim as that of the Haar measure. But before doing so, we need to recall basic properties of Macdonald measures and polynomials of general type \cite{MacdonaldNew}. 
These polynomials are \emph{not} labeled by finite Dynkin diagram; rather, they correspond to (possibly non-reduced) affine root systems $S$.
A reduced affine root system is the usual affine root system underlying (un)twisted affine Lie algebras as discussed by Kac \cite{Kac}.
A non-reduced root system is a root system where ${\vec \alpha}$ and $2{\vec \alpha}$ can simultaneously be roots;
a non-reduced affine root systems $S$ is  labeled by a pair of reduced affine root systems $(X,Y)$ 
where $X=\{{\vec \alpha} \in S :  {\vec \alpha}/2 \not\in S\}$ , $Y=\{{\vec \alpha} \in S :  2{\vec \alpha} \not\in S\}$. 
These  are tabulated in Table~\ref{types}.
\begin{table}[ht]
Reduced:
 \[
\begin{array}{r|cccccccccccccccccccccc}
\text{Kac}			& X^{(1)}_\ell	& A_{2\ell-1}^{(2)}&A_{2\ell}^{(2)} & D_{n+1}^{(2)} & E_6^{(2)} & D_4^{(3)}\\[.3em]
\text{Macdonald}  	& X_\ell 		& B_\ell^\vee        & BC_\ell           & C_\ell^\vee       & F_4^\vee & G_2^\vee
\end{array}
\]
Non-reduced:
 \[
(BC_\ell,C_\ell),\quad
(C_\ell^\vee,BC_\ell),\quad
(B_\ell,B_\ell^\vee),\quad
(C_\ell^\vee,C_\ell).
\]
\caption{List of (possibly-non-reduced) affine root systems\label{types}}
\end{table}

The measure for the Macdonald polynomials of various types is given in \cite{MacdonaldNew}, (5.1.28).  For  simply-laced  $X_\ell$, it was given in \eqref{simplylacedmacdonaldmeasure}. 
For non-simply-laced $X=G$, it is \begin{equation}
\int [\ud {\vec x}]_{G,q,t,t'}=
\int [\ud {\vec x}]_{G} \prod_{n=0}^\infty \prod_{{\vec \alpha}\in \Delta_s} \frac{1-q^{n+1} x^{\vec \alpha}}{1-t q^n x^{\vec \alpha}}
\prod_{{\vec \alpha}\in \Delta_l} \frac{1-q^{(n+1)} x^{\vec \alpha}}{1-t' q^{n} x^{\vec \alpha}}
\end{equation} where $t,t'$ for the short roots and the long roots can be different.

The measure for type $G^\vee$ is given by \begin{equation}
\int [\ud {\vec x}]_{G^\vee,q,t,t'}=
\int [\ud {\vec x}]_{G^\vee} \prod_{n=0}^\infty \prod_{{\vec \alpha}\in \Delta_s} \frac{1-q^{n+1} x^{\vec \alpha}}{1-t q^n x^{\vec \alpha}}
\prod_{{\vec \alpha}\in \Delta_l} \frac{1-q^{(n+1)d} x^{\vec \alpha}}{1-t' q^{nd} x^{\vec \alpha}}
\end{equation} where $\Delta_{s,l}$ are short and long roots for $G^\vee$ (as a finite root system).
Note that we have $q^d$ for the long roots instead of $q$, where $d=2$ for $G^\vee=B,C,F$ and $d=3$ for $G^\vee=G_2$.

The Macdonald polynomials associated to any non-reduced affine root system can be obtained by restricting the parameters of the Macdonald polynomials of type $(C^\vee_\ell,C_\ell)$, which are also called the Koornwinder polynomials. It has the measure 
\def\uq{\check q}
\def\ut{\check t}
\begin{equation}
\int [\ud {\vec x}]_\text{Koornwinder}=
\int [\ud {\vec x}]_{C_\ell} \PE \left[\sum_{{\vec \alpha}\in\Delta_s} \frac{\ut-\uq}{1-\uq}x^{\vec \alpha} + \sum_{{\vec \alpha}\in \Delta_l}\left(\frac{u_1+u_2+u_3+u_4}{1-\uq}x^{{\vec \alpha}/2}+\frac{-\uq}{1-\uq}x^{{\vec \alpha}}\right) \right]\label{Koornwinder}
\end{equation} where $(\uq,\ut,u_1,u_2,u_3,u_4)$ are parameters.

\subsubsection{Vector multiplet measures}\label{vectmes}
In this section, we show that the orthogonal polynomial under the twisted vector multiplet measure is given by \begin{equation}
\tilde K_\text{full}(\vec x)\tilde P_{\vec\lambda}(\vec x),
\end{equation} where $\tilde K_\text{full}(\vec x)$ is a prefactor given in \eqref{tK1}, \eqref{tK2} and $\tilde P_{\vec\lambda}(\vec x)$  is the Macdonald polynomials of type listed in Table~\ref{macdonald-table}. As always, the discussion is split into two cases, depending on whether $\Gamma=A_{2\ell}$.

\paragraph{For $\Gamma\neq A_{2\ell}$:} 
Consider the two-parameter vector multiplet measure for simply-laced $\Gamma\neq A_{2\ell}$ and the $\bZ_d$ twist $\sigma$, as in Sec.~\ref{haar}.  Using $r'$, $r''$ such that $r=r'+r''$ is the rank of $G$ and $r'+dr''$ is the rank of $\Gamma$, 
we obtain 
\begin{equation}
 \PE\left[\frac{-q-t}{1-q}\right]^{r'}
\PE\left[ \frac{-q^d-t^d}{1-q^d}\right]^{r''}
\int [\ud {\vec x}]_{G} \prod_{{\vec \alpha}\in \Delta_s} \PE\left[\frac{-q-t}{1-q}x^{\vec \alpha}\right] 
\prod_{{\vec \alpha}\in \Delta_l} \PE\left[\frac{-q^d-t^d}{1-q^d}x^{\vec \alpha}\right] 
\end{equation} which is 
\begin{equation}
\PE \left[\frac{-q+t}{1-q}\right]^{r'}
\PE \left[\frac{-q^d+t^d}{1-q^d}\right]^{r''}
\int [\ud {\vec x}]_{S,q,t,t'=t^d} 
\tK_\text{full}(\vec x)^{-2}
\end{equation} where \begin{equation}
\tK_\text{full}(\vec x)=
\PE\left[\frac{t}{1-q}\right]^{r'}
\PE\left[\frac{t^d}{1-q^d}\right]^{r''}
\prod_{{\vec \alpha}\in \Delta_s}\PE\left[\frac{t}{1-q}x^{\vec \alpha}\right]
\prod_{{\vec \alpha}\in \Delta_l}\PE\left[\frac{t^d}{1-q^d}x^{\vec \alpha}\right].\label{tK1}
\end{equation}

\begin{table}
\[
\begin{array}{c||cccc|c}
\Gamma & A_{2\ell-1} &  D_{n+1} & D_4 &  E_6 & A_{2\ell} \\
\hline
d & 2 & 2 &  3 & 2   & 2\\
\hline
S & C_\ell^\vee & B_\ell^\vee & G_2^\vee & F_4^\vee   & (C_\ell^\vee,C_\ell) \\
\hline
G & B_\ell & C_\ell & G_2 & F_4 & C_\ell
\end{array}
\]
\caption{Types of the Macdonald polynomials appearing in the vector multiplet measure. Upon 1-parameter specialization $t=q$, we get the Schur functions of type $G$. \label{macdonald-table}}
\end{table}

\paragraph{For $\Gamma\neq A_{2\ell}$:} 
The vector multiplet measure in this case is 
\begin{multline} \label{vpletmeasA2n}
\PE\left[\frac{-q-t}{1-q}\right]^{r'} \PE\left[\frac{-q^d-t^d}{1-q^d}\right]^{r''}  \int [\ud {\vec x}]_{C_\ell}\times \\
 \PE\left[\sum_{{\vec \alpha}\in\Delta_s}\frac{-q^2-t^2}{1-q^2} x^{\vec \alpha} +\sum_{{\vec \alpha}\in\Delta_l}\left(\frac{-q^2-t^2}{1-q^2} x^{{\vec \alpha}/2}+\frac{-uq-ut}{1-q} x^{{\vec \alpha}/2}\right)  \right]_{u=-1}. 
\end{multline}
The first factor in the second line can be rewritten as \begin{equation}
 \PE\left[\frac{-q^2-t^2}{1-q^2} x^{\vec \alpha}\right] 
 = \PE\left[\frac{-\uq+\ut}{1-\uq} x^{\vec \alpha}\right] \PE\left[\frac{-\ut}{1-\uq} x^{\vec \alpha}\right]^2
\end{equation} and the second factor can be massaged as  \begin{multline}
\PE\left[ \left(\frac{-q^2-t^2}{1-q^2} +\frac{-uq-ut}{1-q} \right)x^{{\vec \alpha}/2}  \right]_{u=-1}\\
= \PE\left[ \left(\frac{-q^2+t^2}{1-q^2} +\frac{-uq+ut}{1-q} \right) x^{{\vec \alpha}/2}  \right]_{u=-1}
\PE\left[ \left(\frac{-t^2}{1-q^2}+\frac{-ut}{1-q} \right) x^{{\vec \alpha}/2}\right]^2_{u=-1}.
\end{multline} This last expression can be further rewritten as 
 \begin{multline}
\PE\left[ \left(\frac{-q^2+t^2}{1-q^2} +\frac{-uq+ut}{1-q} \right) x^{{\vec \alpha}/2}  \right]_{u=-1} \\
= \PE \left[\frac{u_1+u_2+u_3+u_4}{1-\uq}x^{{\vec \alpha}/2}+\frac{-\uq}{1-\uq}x^{{\vec \alpha}}\right]_{(\uq,\ut;u_1,u_2,u_3,u_4)=(q^2,t^2;-tq,-t,q,t^2)}.
\end{multline}

Combining all this, we find that the twisted vector multiplet measure equals \begin{equation}
\PE\left[\frac{-q+t}{1-q}\right]^{r'} \PE\left[\frac{-q^d+t^d}{1-q^d}\right]^{r''}
\int [\ud {\vec x}]_\text{Koornwinder} \tilde K_\text{full}(\vec x)^{-2}
\end{equation} where \begin{equation}
[\ud {\vec x}]_\text{Koornwinder}=[\ud {\vec x}]_{(C_\ell^\vee,C_\ell),(\uq,\ut;u_1,u_2,u_3,u_4)=(q^2,t^2;-tq,-t,q,t^2)}
\end{equation} is 
the Koornwinder measure \eqref{Koornwinder}, with the indicated choice of the parameters, and \begin{equation}
\tilde K_\text{full}(\vec x)=
\PE\left[\frac{t^2}{1-q^2}\right]^{r''}
\prod_{{\vec \alpha}\in \Delta_s}\PE\left[\frac{t}{1-q}x^{\vec \alpha}\right]
\prod_{{\vec \alpha}\in \Delta_l}\PE\left[\frac{ut}{1-q}x^{{\vec \alpha}/2}+\frac{t^2}{1-q^2}x^{{\vec \alpha}/2}\right]_{u=-1}.
\label{tK2}
\end{equation}

Let us make some observations. 
First,  the vector multiplet measure for $G=C_\ell$ obtained from $\Gamma=A_{2\ell}$ is different from the one  for $G=C_\ell$ obtained from $\Gamma=D_{n+1}$. From the point of view of the 5d maximally-supersymmetric Yang-Mills, this should stem from the difference in the discrete theta angle taking value in $\pi_4(G)=\bZ_2$.
Second, the measure in the one-parameter version can easily be obtained by setting $t=q$ in the formulas above. The vector multiplet measures for the two ways of obtaining $G=C_\ell$ still differ.

\section{Contributions from the matter fields}\label{hyper}
Let us now move on to the study of the 4d theory corresponding to a general Riemann surface $\cC_2$ with punctures. 
Any Riemann surface can be decomposed to cylinders and three-punctured spheres, and we already discussed the contribution from the cylinder in the last section. Therefore, we have to study the superconformal indices of the 4d theories corresponding to three-punctured spheres, which we call as matter fields. 

\subsection{Without outer-automorphism twists}
A puncture for the 6d theory of type $G=A_{n-1}$ is specified by a Young diagram ${\vec \Lambda}$ with $n$ boxes, or equivalently an embedding \begin{equation}
{\vec \Lambda}:\SU(2)\to G. \label{embedding}
\end{equation}
The partition ${\vec N}=[l_1,l_2,...]$ means that under the corresponding ${\vec \Lambda}: \SU(2) \rightarrow \SU(N)$,
the fundamental of $\SU(N)$ decomposes into a direct sum of irreducible representations of $\SU(2)$ of dimension $l_1, l_2, \ldots$ respectively.

The flavor symmetry $G_{\vec \Lambda}$ associated to the puncture is given by the commutant of ${\vec \Lambda}(\SU(2))$ in $G$:
\begin{equation}
 G_{\vec \Lambda} \times {\vec \Lambda}(\SU(2)) \subset G. \label{flavor}
\end{equation}
Thus, the fugacity $a$ of the flavor symmetry of a puncture can naturally be thought of as a fugacity of $\SU(n)$.  The superconformal index of the theory for the three-punctured sphere with punctures ${\Lambda}_{1,2,3}$ is conjectured to have the following form \cite{Gadde:2011uv}: \begin{equation}
\cI = \sum_{\vec \lambda} \frac{\prod_{I=1}^3 K_{{\Lambda}_I}(\vec{a}_I) \underline{P}_{\vec \lambda}(\vec{a}_I t^{{\Lambda}_I})}{K_{\vec \rho} \underline{P}_{\vec \lambda}(t^{\vec \rho})}.\label{3pt-simply-laced}
\end{equation} Here, we made the obvious generalization of the formula in \cite{Gadde:2011uv} to a general simply-laced group $G$, and our $K_{{\vec \Lambda}}(\vec a)$ is different from their $\hat{\mathcal{K}}_{\vec \Lambda}(\vec a)$ by an $a$-independent $(q,t)$-dependent factor.  Let us explain the notations. 
\begin{itemize}

\item $\underline{P}_{\vec \lambda}(\vec x)=N_{\vec \lambda}^{-1/2} P_{\vec \lambda}(\vec x)$ is the normalized Macdonald polynomial orthonormal under the measure \begin{equation}
\frac{(q;q)^r}{(t;q)^r}[\ud {\vec x}]_{G,t,q} = [\ud {\vec x}]_G \PE \left[\frac{-q+t}{1-q}\chi_\text{adj}(\vec x) \right]
\end{equation}  where \begin{itemize}
	\item $(a;q)$ is the $q$-Pochhammer symbol \begin{equation}
	(a;q)=\prod_{i=0}^\infty (1-aq^i).
	\end{equation}
	\item   $r$ is the rank of $G$. 
	\item We follow the convention in the mathematics literature by reserving the letter $P_{\vec \lambda}(\vec x)$ to be the unnormalized orthogonal Macdonald polynomial satisfying \begin{equation}
P_{\vec \lambda}(\vec x)=\chi_{\vec \lambda}(\vec x)+\sum_{\vec \mu<\vec \lambda} f_{\vec \lambda}(q,t) \chi_{\vec\mu}(\vec x) , 
\end{equation} where the sum is over $\vec \mu$ such that ${\vec \lambda}-\vec \mu$ is a sum of positive number of simple roots.  
The series expansions of $P_{\vec \lambda}(\vec x)$ and $N_{\vec \lambda}$ in terms of $q$ and $t$ are known to have integer coefficients.  This guarantees that the series expansion of the formula \eqref{3pt-simply-laced} has integer coefficients, as it should be for a superconformal index.
\end{itemize}

\item The argument of the Macdonald polynomials in the numerator is given by $\vec{a} t^{\vec \Lambda}$. Here, $\vec{a} \in G_{\vec \Lambda}$ and $t$ is now thought of as the element $t\in \SU(2)$ given by $\diag(t^{1/2},t^{-1/2})$. Then $t^{\vec \Lambda}$ is defined to be the image of $t\in \SU(2)$ via the map \eqref{embedding} in ${\vec \Lambda}(\SU(2))$. Then $a t^{\vec \Lambda}$ is naturally an element of $G$ via the embedding \eqref{flavor}.

The physical reason behind the factor $t^{\vec\Lambda}$ is as follows. The puncture of type $\vec\Lambda$, as a boundary condition of 5d maximally supersymmetric Yang-Mills of gauge group $G$, is described by giving the profile \cite{Gaiotto:2008ak,Chacaltana:2012zy}\begin{equation}
\Phi_i(s)=\vec\Lambda(\tau_i)/s \ , \label{profile}
\end{equation} where $\tau_i$ are the standard generators of $\SU(2)$, $s$ is the distance to the boundary, and $\Phi_{1,2,3}$ are three scalars of the theory, transforming as a triplet under the $\SU(2)_R\subset \SO(5)_R$ symmetry of the 5d Yang-Mills.  This profile is only invariant under the diagonal subgroup $\SU(2)_R\times \vec\Lambda(\SU(2))$, which becomes the $\SU(2)_R$ symmetry of the resulting 4d theory. From the definition of the $\cN=2$ superconformal index in \eqref{n2index}, it is then clear that $t$ also appears as a fugacity of the $G$-flavor symmetry.

\item  In the denominator,  $t^{\vec \rho}$ is defined using the principal embedding \begin{equation}
{\vec \rho}:\SU(2)\to G
\end{equation} and in particular \begin{equation}
t^{\vec \rho}=\diag(t^{(n-1)/2},t^{(n-3)/2},\ldots, t^{(1-n)/2})
\end{equation} for $G=\SU(n)$. The exponents are given by the coordinates of the Weyl vector. 

\item  $K_{\vec \Lambda}(\vec a)$ is a prefactor independent of ${\vec \lambda}$, determined as follows.  The embedding \eqref{embedding}, \eqref{flavor} induces the decomposition \begin{equation}
\fg = \bigoplus_j R_j \otimes V_j , \label{flavor-decomp} 
\end{equation} where $\fg$ is the Lie algebra of $G$, $R_j$ is a representation of $G_{\vec \Lambda}$ and $V_{2j+1}$ is the spin-$j$ irreducible representation  of $\SU(2)$. 
Note that $\bigoplus_j R_j$ gives the decomposition of the Slodowy slice \cite{Chacaltana:2012zy}. Each component in the slice give rise to a plethystic exponential, giving \begin{equation}
K_{\vec \Lambda}(\vec a)=\PE \left[\sum_j\frac{t^{j+1}}{1-q} \tr_{R_j}(\vec a)\right].  \label{K}
\end{equation}
\begin{itemize}
	\item For the full puncture, the corresponding embedding ${\Lambda}_\text{full}:\SU(2)\to G$ is the zero map. Then  	\begin{equation}
K_\text{full}(\vec a)=\PE \left[  \frac{t}{1-q} \chi_\text{adj}(\vec a) \right]. \label{Kfull}
\end{equation}
Note that $R_{\vec \lambda}(\vec a)\equiv K_\text{full}(\vec a)\underline{P}_{\vec \lambda}(\vec a)$ satisfies the orthonormality condition under the vector multiplet measure \eqref{measure2} thanks to the equality \eqref{ooo}: \begin{equation}
\int [\ud {\vec x}]_G \PE \left[\frac{-q-t}{1-q}\chi_\text{adj}(\vec x)\right] R_{\vec \lambda}(\vec x)R_\mu(x^{-1}) 
= \delta_{{\vec \lambda}{\vec \mu}}.\label{ortho}
\end{equation}

	\item For the zero puncture, the corresponding embedding is given by ${\vec \rho}:\SU(2)\to G$ is the principal embedding. Then the decomposition \eqref{flavor-decomp} is \begin{equation}
\fg = \oplus_{i=1}^r V_{d_i-1},
\end{equation}  
where the integers $d_{1,\ldots,r}$ are the degrees of invariants of $G$, so that $d_i=i+1$ when $G=A_\ell$. 
Then \begin{equation}
K_{\vec \rho}=\PE \left[\sum_i\frac{t^{d_i}}{1-q} \right] = \prod_{i=1}^r (t^{d_i};q)^{-1}. 
\end{equation} This also appears in the denominator.
\end{itemize}

\end{itemize}

 From the orthonormality \eqref{ortho}, the superconformal index for the Riemann surface with genus $g$ and $s$ punctures ${\Lambda}_{I=1,\ldots,s}$ can easily be computed: \begin{equation}
  \cI = \sum_{\vec \lambda} \frac{\prod_{I=1}^s  K_{{\Lambda}_I}(\vec{a}_I) \underline{P}_{\vec \lambda}(\vec{a}_I t^{\vec\Lambda_I})}{[K_{\vec \rho} \underline{P}_{\vec \lambda}(t^{\vec \rho})]^{2(g-1)+s}}.\label{general-simply-laced}
\end{equation}
This expression also makes it clear that adding a puncture of type ${\vec \rho}$, which corresponds to the absence of any puncture, makes no change in the partition function.

\subsection{With outer-automorphism twists} \label{sec:withtwists}
Let us generalize the results \eqref{3pt-simply-laced} and \eqref{general-simply-laced} in the previous section to the type-$A$ theories with twists. 
Here we will first state the final formula. 
Ample pieces of evidence are given later, by considering various three-punctured spheres giving free hypermultiplets, and the Argyres-Seiberg duality.

Take the 4d theory arising from a  sphere of the $A_{N-1}$ theory with three punctures of type ${\Lambda}_{1,2,3}$ respectively. 
We find that the superconformal index with the $\bZ_2$ twist is given by \begin{equation}
 \tilde{\cI} = \sum_{\vec \lambda} \frac{\prod_{I=1}^3 \tilde K_{{\Lambda}_I}(\vec{a}_I) \underline{\tP}_{\vec \lambda}(\vec{a}_I t^{{\Lambda}_I})}{\tilde K_{\vec \rho} \underline{\tP}_{\vec \lambda}(t^{\vec \rho})}.\label{generaltwistedformula}
\end{equation}
The symbols are explained below: \begin{itemize}
\item $\underline{\tP}_{\vec \lambda} = \tilde{N}_{\vec \lambda}^{-1/2}\tP_{\vec \lambda}$ are the normalized Macdonald polynomials, orthonormal under the measure  
\begin{equation}
 \frac{(q;q)^{r'}}{(t;q)^{r'}} \frac{(q^2;q^2)^{r''}}{(t^2;q^2)^{r''}} [\ud {\vec x}]_\text{Macdonald}.
\end{equation}
Here, $[\ud {\vec x}]_\text{Macdonald}$ is the Macdonald measure of the type and the parameters as defined in Sec.~\ref{vectmes}. Also, as given in that section, $r'$ and $2r''$ are the dimensions of the $\bZ_2$ invariant and non-invariant parts of the Cartan. Concretely, we have 
\bea
(r',r'')=(1,l-1) \quad \text{if $N=2l$}, \qquad (r',r'')=(0,l) \quad  \text{if $N=2l+1$}.
\eea
Here $\tP_{\vec \lambda}$  are the unnormalized Macdonald polynomials.  When $t=q$, they reduce to characters of $B_l$ for $N=2l$ and of $C_l$ for $N=2l+1$.  The way to calculate them is described in Appendix~\ref{computation}.

We use the convention that 
\bea
\begin{array}{rcllcl}
\chi_\text{fund}^{C_l}(\vec z)&=&E_1(\vec z)~, \quad &\chi_\text{vec}^{B_l}(\vec z)&=&1+E_1(\vec z)~,\nn \\
\chi_\text{adj}^{B_l}(\vec z) &=& l +E_1(\vec z)+E_2(\vec z)~, \quad & \chi_\text{spin}^{B_l}(\vec z) &=& \widehat{E}_l(\vec z)~.
\end{array}
\eea
Here and in the following, we use the abbreviations
\bea
E_1(\vec z) &=\sum_{i=1}^l z_i^2 + z_i^{-2}~, \nn \\
E_2(\vec z) &=\sum_{1 \leq i < j \leq l} z_i^2 z_j^2 +  z_i^2 z_j^{-2}+  z_i^{-2}z_j^2 +  z_i^{-2} z_j^{-2}~, \nn \\
\widehat{E}_l(\vec z) &= \sum_{\epsilon_i = \pm 1} z_1^{\epsilon_1} \ldots z_l^{\epsilon_l}~.
\eea
\item In the untwisted case we had  $P_{\vec \lambda}(\vec a)$ in both the denominator and the numerator, where
\begin{equation}
{\vec a}t^{{\vec \Lambda}} = \diag(a_1 t^{{\Lambda}_1},a_2 t^{{\Lambda}_2},\ldots, a_{N} t^{{\Lambda}_N}).
\end{equation} 

Recall that the powers of $t$ comes from the necessity to preserve the boundary profile \eqref{profile}. 
Here we need to make the action of the outer automorphism $\cP_\sigma$ compatible with it. 
Namely, we pick an inner action $B_\sigma$ such that \begin{equation}
B_\sigma \cP_\sigma (\vec{\Lambda}(\tau_i) ) B_\sigma^{-1} =  \vec{\Lambda}(\tau_i) \ ,
 \label{invariance}
\end{equation} where $\tau_i$ are the standard generators of $\SU(2)$.
For $\tau_3$, this can be achieved by making ${\Lambda}_{i}=-{\Lambda}_{N+1-i}$ (for $i=1, \ldots, N$) by reordering. 
For $\tau_\pm=\tau_1\pm i\tau_2$, the equation above can be solved by a diagonal $B_\sigma$ with $\pm1$, $\pm i$ as entries.

Therefore, the rule to obtain the argument of the orthogonal polynomials becomes the following: 
\ben
\item Restrict the flavor fugacities so that we have
\bea
\text{for $N=2l$,} \quad a_i &= a^{-1}_{N+1-i}~, \quad \forall~ i=1, \ldots, l~; \nn \\
\text{for $N=2l+1$,} \quad a_i &= a^{-1}_{N+1-i}~,\quad a_{l+1}=1 \quad \forall~i=1, \ldots, l~. \label{restrict}
\eea
\item Multiplies with factors of $\pm1$, $\pm i$ coming from $B_\sigma$.
\item Select half of the elements from the tuple, one from each pair exchanged by $\bZ_2$ action.  
As a convention we choose those whose powers in $t$ are non-negative.
\een
For example, \begin{itemize}
	\item For ${\vec \Lambda}_\text{full}=[1^N]$, i.e.~ the full puncture, the profile $\vec\Lambda(\tau_i)=0$ automatically satisfies \eqref{invariance}. So we have
	\bea
	{\vec a}t^{\vec \Lambda} =(a_1,a_2,\ldots,a_l)~, &\qquad l = \lfloor N/2 \rfloor~.
	\eea
	\item For $\vec\rho={\vec \Lambda}=[N]$, i.e.~the null puncture, the profile \begin{equation}
	\vec\Lambda(\tau_+) = \begin{cases}
	c_0 e_{-1,1} + \sum_{s=1}^{l-1} c_s (e_{-s-1,-s} +e_{s,s+1}), & \qquad N=2l \\
	 \sum_{s=0}^{l-1} c_s (e_{-s-1,-s} +e_{s,s+1}), & \qquad N=2l+1 \\
	\end{cases}
	\end{equation} is invariant under the action $\cP_\sigma$ given in \eqref{explicitZ2actionAodd}, \eqref{explicitZ2actionAeven}. 
	For example,  
	 \begin{equation}
	{\vec \Lambda}(\tau_+)=\begin{pmatrix}
0 & \bullet  &  0 & 0 \\
0 & 0  & \bullet &  0 \\
0 & 0  & 0 &\bullet \\
0 & 0  & 0 & 0
\end{pmatrix}  
	\end{equation} for $N=4$; see the right diagram in Fig.~\ref{action}.
	We have
	\bea
	{\vec a}t^{\vec \Lambda} = t^{\vec \rho} = \begin{cases} (t^{l-1/2},t^{l-3/2},\ldots, t^{1/2})~, &\qquad N=2l \\
	    (t^{l},t^{l-1},\ldots, t)	~, &\qquad N=2l+1~. \label{trho}
	\end{cases}   
	\eea
	
	\item For ${\vec \Lambda}_\text{simple}=[N-1,1]$, i.e.~the simple puncture, the profile \begin{equation}
	\vec\Lambda(\tau_+) = \begin{cases}
	c'_0 (e_{-2,1} + e_{-1,2}) + \sum_{s=1}^{l-1} c'_s (e_{-s-1,-s} +e_{s,s+1}), & \qquad N=2l \\
	c'_0 e_{-1,1} + \sum_{s=1}^{l-1} c'_s (e_{-s-1,-s} +e_{s,s+1}), & \qquad N=2l+1 
	\end{cases}
	\end{equation} is \emph{not} invariant under the action $\cP_\sigma$ given in \eqref{explicitZ2actionAeven}, \eqref{explicitZ2actionAodd}. 
	For example, \begin{equation}
	{\vec \Lambda}(\tau_+)=\begin{pmatrix}
0 & \bullet  & \bullet & 0 \\
0 & 0  & 0 &  \bullet \\
0 & 0  & 0 &\bullet \\
0 & 0  & 0 & 0
\end{pmatrix} 
	\end{equation} for $N=4$, and compare it with the right diagram of Fig.~\ref{action}.
	This non-invariance can be canceled by \begin{equation}
	B_\sigma =\begin{cases}
	\diag(\underbrace{i,\ldots,i}_l,\underbrace{-i,\ldots,-i}_l), & \qquad N=2l\\
	\diag(\underbrace{i,\ldots,i}_l,1,\underbrace{-i,\ldots,-i}_l), & \qquad N=2l+1.
	\end{cases}
	\end{equation} Then we have
	\bea
	t^{\vec \Lambda_\text{simple}} =  \begin{cases} (it^{l-1},it^{l-2},\ldots, i)~, &\qquad N=2l \\
	    (it^{l-1/2},it^{l-1/2},\ldots, it^{1/2})	~, &\qquad N=2l+1~. \label{tsimple}
	\end{cases}   
	\eea
	
	More examples are discussed in the following sections.
	\end{itemize}
\item In the untwisted case, the prefactor $K_{\vec \Lambda}(\vec a)$ is given by the plethystic exponential \eqref{K}, i.e.~by products of infinite products of the form 
\bea
K_{\vec \Lambda}(\vec a) =\prod_{s}\prod_{n=0}\frac{1}{1-z_{\Lambda,s} (\vec a,q,t) q^{n}} \ , 
\eea
where $z_s(q,t,a)$ are {\it monomials} of $q$, $t$ and the fugacities $a$. Each choice of the index $s$ corresponds to a basis of the Slodowy slice, and generates the Fock space of a Kaluza-Klein tower with the fugacity $z_s$ and the mode number $n$.  The $\bZ_2$ twist assigns a $\pm$ sign to each of the factors, resulting in \begin{equation} \label{twistedprefacgen}
\tilde K_{\vec \Lambda}(\vec a) =\prod_{s}\prod_{n=0}\frac{1}{1\pm z_{\Lambda,s}(\tilde{\vec a},q,t) q^{n}} \ ,
\end{equation} where $\pm1$ is determined by the action of $B_\sigma \cP_\sigma$ on the Slodowy slice. 
For example, \begin{itemize}
	\item when ${\vec \Lambda}=[1^N]$, the untwisted prefactor $K_{\vec \Lambda}(\vec a)$  was the conversion factor between the vector multiplet measure and the Macdonald measure, as discussed in \eqref{Kfull}, \eqref{ortho}. 
	The twisted prefactor is also the conversion factor between them:
	\bea \label{tKfull}
	\tK_\text{full}(\vec z)= \begin{cases} 
	\frac{(-t;q)}{ (t;q)^l (-t;q)^l } \PE \left[\frac{t}{1-q}E_1(\vec z) + \frac{t^2}{1-q^2} E_2(\vec z) \right] ~, &~ N=2l \\
	\frac{(-t,q)}{(-t;q)^{l+1} (t;q)^{l}} \PE \left[\frac{u t}{1-q}E_1(\vec z) + \frac{t}{1-q} E_2(\vec z)+ \frac{t^2}{1-q^2} E_1(\vec z) \right]_{u=-1} ~, &~ N=2l+1~.
	\end{cases}
	\eea
	\item when ${\vec \Lambda}=[N]$, we have \begin{equation}
	\tK_{[N]} = \tK_{{\vec \rho}} =\prod_{i=2}^{N}\frac1{((-t)^i;q)} \ . \label{tKrho}
\end{equation}
	\end{itemize}
	The general rules for the twisted prefactors are given in Appendix~\ref{rule}. More examples can be found in the following sections.
\end{itemize}

Let us now move on to the examples where we can test the formula explained above. 

\subsubsection{Free bifundamental hypermultiplets}
The two-parameter index of a hypermultiplet in the representation $R$ without outer-automorphism twist is given by the formula \begin{equation}
\PE \left[\frac{t^{1/2}}{1-q} \chi_{R \oplus \bar R}(\vec x) \right].
\end{equation}
In particular, for a bifundamental hypermultiplet of $\SU(N)\times \SU(N)$, we have \begin{equation}
\CI^{\text{bifund}}_N (q,t, \vec x, \vec y) = \PE \left[\frac{t^{1/2}}{1-q} \left \{ \chi^{\SU(N)}_\text{fund}(\vec x)\chi^{\SU(N)}_\text{fund}(\vec{y}^{-1})+\chi^{\SU(N)}_\text{fund}(\vec{x}^{-1})\chi^{\SU(N)}_\text{fund}(\vec y) \right\} \right].
\end{equation}

As was the case for vector multiplet contributions studied in Sec.~\ref{vectortwist}, this infinite product comes from the trace over the Fock space generated by the raising operators \begin{equation}
v_{i,j;n},\quad w_{i,j;n}
\end{equation} which have the charge $x_i/y_j t^{1/2}q^n$ and $y_j/x_i t^{1/2}q^n$, respectively. 
From the charge assignments, it is clear that the $\bZ_2$ operation just exchanges the modes 
$v_{i,j;n}$ and $w_{i,j;n}$.
The superpotential term contains a term of the form $\sum_i e_{i,i} v_{i,j;n}w_{i,j;n}$ where $e_{i,i}$ denotes a mode in the vector multiplet with the specified gauge indices. This should be invariant under the $\bZ_2$ action, which acts on $e_{i,i}$ via $e_{i,i}\to -e_{-i,-i}$.
Therefore we find that the $\bZ_2$ action on $v_{i,j;n}$ and $w_{i,j;n}$ needs to have  the form 
\begin{equation} \label{bifundmodetwist}
v_{i,j;n}\to w_{i,j,n},\qquad
w_{i,j;n}\to -v_{i,j,n}.
\end{equation}  

To perform the $\bZ_2$ operation inside the trace, one now needs to do the following replacements:
\bea
\text{For $N=2l$,} \quad x_i &= x^{-1}_{N+1-i}~, \quad y_i = y^{-1}_{N+1-i}~, \quad \forall~ i=1, \ldots, l~, \nn \\
\text{For $N=2l+1$,} \quad x_i &= x^{-1}_{N+1-i}~, \quad y_i = y^{-1}_{N+1-i}~, \quad x_{l+1}= y_{l+1} =1 \quad \forall~i=1, \ldots, l~.    \label{replbifund}
\eea
We can therefore obtain the twisted character of the bifundamentals in two steps: (1) performing the replacements \eref{replbifund}, and (2) combining of two modes to get a $\BZ_2$ invariant combination (``doubling'').  The process can be explicitly written down as follows:
\bea
& \chi^{\SU(N)}_\text{fund}(\vec x)\chi^{\SU(N)}_\text{fund}(\vec{y}^{-1})+\chi^{\SU(N)}_\text{fund}(\vec{x}^{-1})\chi^{\SU(N)}_\text{fund}(\vec y) \nn \\
& \overset{\text{Step (1)}}{\longrightarrow} \begin{cases} 2E_1(\vec{x}^{1/2}) E_1(\vec{y}^{1/2})~, &\quad N=2l \\
 2(E_1(\vec{x}^{1/2})+1)( E_1(\vec{y}^{1/2})+1)~, &\quad N=2l+1  \end{cases} \nn \\
& \overset{\text{Step (2)}}{\longrightarrow}  \begin{cases} E_1(\vec x) E_1(\vec y)~, &\quad N=2l \\
 (E_1(\vec x)+1)( E_1(\vec y)+1)~, &\quad N=2l+1~.  \end{cases}
\eea
Thus, the superconformal index is given by 
\bea\label{twistedbifund}
\tI^{\text{bifund}}_N = \begin{cases} \PE \left[ \frac{ut}{1-q} E_1(\vec x)E_1(\vec y) \right]_{u=-1}~, &\quad N=2l  \\ 
\PE \left[ \frac{ut}{1-q} (E_1(\vec x)+1)(E_1(\vec y)+1) \right]_{u=-1}~, &\quad N=2l+1~. 
\end{cases}
\eea
Note that we used fermionic plethystic exponent since we get the minus sign in the index from \eqref{bifundmodetwist} for a paired up mode $v_{i, j; n} w_{i, j; n}$. 

Let us next consider the three-punctured spheres realizing the free bifundamental hypermultiplet, which is a sphere with two full punctures and one simple puncture. 
The arguments of Macdonald polynomials are already discussed, see \eqref{trho} and \eqref{tsimple}. 
The prefactor $\tK_\text{simple}$ is determined to be \begin{align}
\tK_{[2l-1,1]}&=\frac{1}{(t^{2l};q^2)}\prod_{i=1}^{2l-1} \frac{1}{((-t)^i;q)},&
\tK_{[2l,1]}&=\frac{1}{(-t^{2l+1};q^2)}\prod_{i=1}^{2l} \frac{1}{((-t)^i;q)}.
\end{align} The general formula then gives the twisted index \begin{equation}
\tI^{\text{bifund}} = \frac{\tK_\text{full}(\vec x) \tK_\text{full}(\vec y) \tK_\text{simple} }{\tK_{\vec \rho}} \sum_{\vec \lambda} \frac{ 
 \underline{\tP}_{\vec \lambda}(\vec x)
\underline{\tP}_{\vec \lambda}(\vec y)
 \underline{\tP}_{\vec \lambda}(it^{\vec\Lambda_\text{simple}})
}
{ \underline{\tP}_{\vec \lambda}(t^{\vec \rho})}. 
\end{equation} 
where $\tK_\text{full}$, $\tK_{\vec \rho}$  are given by \eref{tKfull}, \eref{tKrho}, respectively.
We checked the agreement of it with \eqref{twistedbifund} for the cases of $N=2, 3, 4, 5, 6, 7$ by expanding both sides in series in $q$, $t$ by Mathematica.
The twisted superconformal index of other three-punctured spheres which give free hypermultiplets are discussed in Appendix~\ref{other}.

\subsubsection{The Argyres-Seiberg duality}

In \cite{Gadde:2010te}, the superconformal index of the $T_3$ theory has been computed using the inverse formula of certain elliptic beta integral \cite{SWinversion}. From the Argyres-Seiberg duality \cite{Argyres:2007cn}, $\SU(3)$ theory with $N_f=6$ is dual to $E_6$ SCFT with $\SU(2)$ flavor subgroup gauged and coupled to a fundamental hypermultiplet. Schematically, the superconformal index of the theory can be written as
\be
 \cI(a, \vec{x}, b, \vec{y}) &=& \int [\ud {\vec z}]_{\SU(3)} ~ \CI^\text{hyp}_{\SU(3)}(a, \vec{x}; \vec{z}) \CI^V_{\SU(3)} (\vec z) \CI^\text{hyp}_{\SU(3)}(b, \vec{y}; {\vec z}) \nn \\
 &=& \int [ds]_{\SU(2)}~ \CI_{E6} (\vec x; \vec y; a, s) \CI^V_{\SU(2)} (s) \CI^\text{hyp}_{\SU(2)} (s; b) \ , 
\ee
where $a, b$ are $\UU(1)$ flavor fugacities and $\vec{x}, \vec{y}, \vec{z}$ are the $\SU(3)$ flavor fugacities. Let's explain the physical meaning of each line briefly.
\bi
\item In the first line, we start from two identical theories, each containing the $\SU(3) \times \SU(3)$ free bifundamentals.  One of the $\SU(3)$ global symmetries from each theory is then gauged and coupled together. The fugacities $\vec z$ correspond to the gauged $\SU(3)$ group and $\CI^V_{\SU(3)} (\vec z)$ is the corresponding vector multiplet index.   
\item In the second line, we decompose the $E_6$ global symmetry into its maximal subgroup $\SU(3)^3$.  We denote by $\vec x$ and $\vec y$ the fugacities of two of these $\SU(3)$.  The remaining $\SU(3)$ global symmetry is decomposed further into its subgroup $\SU(2) \times \UU(1)$, whose fugacities are denoted by $s$ and $a$ respectively.  This $\SU(2)$ is then gauged and coupled to the theory with a free bifundamental in $\SU(2) \times \UU(1)$, whose fugacities are denoted by $s$ and $b$ respectively. Here $\CI^V_{\SU(2)} (s)$ is the vector multiplet index corresponding to the gauged $\SU(2)$ group.
\ei


The same kind of relation should hold for the twisted index due to the duality. Even though there is no analogous inversion formula for the twisted index, we can use the TQFT structure of the index to write down a natural conjecture for the index \cite{Gadde:2011ik, Gadde:2011uv}. In this section, we will conjecture such a twisted index for the $E_6$ SCFT and then verify if against the Argyres-Seiberg duality. 

Let's first start with the case with the weakly coupled frame. The vector multiplet twisted index for the $\SU(3)$ theory is given by the integrand of \eref{vpletmeasA2n} with $(r'=0, r''=1)$:
\be
{\tI}^V (x) = \PE \left[ \frac{-q^2 - t^2}{1-q^2} \left( x^2 + \frac{1}{x^2} + 1\right) + \frac{- uq - ut}{1-q} \left( x^2 + \frac{1}{x^2} \right)\right]_{u=-1} , 
\ee
and the hypermultiplet contribution is given by \eref{twistedbifund}:
\be
{\tI}^{\text{hyp}} (z; x) = \PE \left[ \frac{ut}{1-q^2} \left( x^2 + \frac{1}{x^2} + 1 \right) \left( z^2 + \frac{1}{z^2} + 1 \right) \right]_{u=-1} \ . 
\ee
The twisted index for the $\SU(3)$ theory with $N_f=6$ is now given by
\be
\tI_{weak}(z, y) = \oint_{|x|=1} \frac{dx}{2\pi i x} \widetilde{\Delta}_{A_2}(x) {\tI}^V (x) {\tI}^\text{hyp}(z; x^{-1}) {\tI}^\text{hyp}(y; x) .
\ee
where 
\bea
\tilde{\Delta}_{A_2}(x) = \Delta_{C_1}(x) = \frac{1}{2}(1-x^4)(1-x^{-4}) \label{HaarC1}
\eea
is the Haar measure of $C_1$. 

Now, in the dual strongly coupled frame, the twisted index should be written as
\be
{\tI}_{strong} (z; y) = \oint_{|x|=1} \frac{dx}{2\pi i x} \widetilde{\Delta}(x) {\tI}^V_{\SU(2)} (x) {\tI}^\text{hyp}_{\SU(2)} (x) {\tI}_{E_6} (x, y, z) . 
\ee
Here the important thing is that the twisted Haar measure $\widetilde{\Delta}(x)$ is not that of $C_1$ as one would naively expect. Note that we need to decompose $3$ of $\SU(3)$ into $2\oplus1$ of $\SU(2)$ which should be compatible with the $\s$ action.  Since $\sigma$ acts on $\SU(3)$ by sending three indices $1,0,-1$ to $-1,0,1$ (see \eref{restrict}), the $\SU(2)$ subgroup compatible with this action is the subgroup of matrices of the form:
\be
\left(
\begin{array}{ccc}
 a & 0 & e_{1, -1}   \\
 0 & 0 & 0  \\
 e_{-1, 1} &  0 & -a
\end{array}
\right) . 
\ee
Since $\sigma$ maps $e_{i, j}$ to $- e_{-j, -i}$, the $\SU(2)$ subgroup given above keeps only the modes $e_{1,-1}$ and $e_{-1,1}$ out of the six roots of $\SU(3)$. The twisted Haar measure is thus given by
\be
 \widetilde{\Delta}( x) = \half (1+x^2) \left(1+\frac{1}{x^2} \right). \label{twistedhaar}
\ee 
Note the sign difference from the Haar measure \eref{HaarC1} of $C_1$; the plus signs in \eref{twistedhaar} are due to the minus sign in the action of $\cP_\sigma$ in \eref{A2nCharges}. 

The vector multiplet contribution is not also the same as the twisted $\SU(2)$ or $\SU(3)$. Since the gauge group is coming from gauging the $\SU(2) \subset SU(3)$, we find 
\be
{\tI}^V_{\SU(2)} (x) &=& \PE\left[ \frac{-q-t}{1-q} \right] \PE \left[ \frac{- u q - ut}{1-q} \left(x^2 + \frac{1}{x^2} \right)\right]_{u=-1} . 
\ee
The first term comes from the Cartan of $\SU(2)$ and the second term comes from the states built out of roots $e_{\pm 1, \mp 1}$. 
The hypermultiplet contribution is simply given by \eref{twistedbifund}:
\be
{\tI}^\text{hyp}_{\SU(2)} (x) &=& \PE \left[ \frac{ut}{1-q^2} \left( x^2 + \frac{1}{x^2} \right) \right]_{u=-1} .
\ee

From the TQFT structure, suppose the twisted index of Minahan-Nemeschansky $E_6$ theory is given by a 3-point function for the 3 full-punctures of the form 
\be
{\tI}_{E_6}( x, y, z) = {\tK}_{\text{full}}(x) {\tK}_{\text{full}}(y) {\tK}_{\text{full}}(z) \sum_\l \frac{\underline{\tP}_\l (x) \underline{\tP}_\l(y) \underline{\tP}_\l (z) }{\tilde{K}_{\vec \rho} \underline{\tP}_\lambda (t^\rho)} . 
\ee
The prefactors ${\tK}_{\text{full}}={\tK}_{[1,1,1]}$ and ${\tK}_{\vec \rho}$ are given by \eref{tKfull} and \eref{tKrho} respectively, and $\tilde P_\lambda (x)$ is the Askey-Wilson polynomial or the Macdonald polynomial of type $(C_1^\vee, C_1)$.

The strongly coupled frame answer is given by
\be
{\tI}_{strong} &=& \int \frac{dx}{2\pi i x} \tilde{\Delta}(\vec x) {\tI}_{E_6} (x, y, z) {\tI}^V_{\SU(2)}(x) {\tI}^\text{hyp}_{\SU(2)}(x) \ . 
\ee
We find 
\be
{\tI}_{strong}  ={\tI}_{weak} 
\ee
as expected from the Argyres-Seiberg duality. Of course it can also be expressed by the general formula \eqref{generaltwistedformula} applied to the four-punctured sphere with two full punctures and two simple punctures.

\section*{Acknowledgments}
NM and JS are grateful for the hospitality of the 2012 Simons Workshop in Mathematics and Physics at the Simons Center for Geometry and Physics during the initial stage of this project. NM would like to thank Dieter L\"ust and Stefan Hohenegger for hospitality at CERN during the completion of the project.  The work of NM is supported by a research grant of the Max Planck Society. The work of JS is supported by DOE-FG03-97ER40546. The work of YT is partially supported by World Premier International Research Center Initiative (WPI Initiative), MEXT, Japan.

\appendix

\section{Study of the 2d propagator}\label{propagator}
Let us consider the vector multiplet measure from the point of view in \cite{Tachikawa:2012wi}. 
The 4d theory obtained by compactifying the 6d theory of type $\Gamma$ on a cylinder with finite area is a $\cN=2$ non-linear sigma model whose target space is $T^*\Gamma_\bC$. 
Here and in the following we use $\Gamma$ to denote the group.
Its partition function on $S^3\times S^1$ boils down essentially to a partition function of a quantum mechanics on the group manifold $\Gamma$.
By the theorem of Peter-Weyl, the wave functions on the group manifold $\Gamma$ has the irreducible decomposition under the $\Gamma\times\Gamma$ action as \begin{equation}
\text{functions on $\Gamma$} \simeq \bigoplus_{\vec \lambda} R_{\vec \lambda}\otimes \bar R_{{\vec \lambda}},
\end{equation} where $R_{\vec \lambda}$ is the irrep with the highest weight ${\vec \lambda}$. Then the trace of the chemical potentials $(\vec x,y)$ for $\Gamma\times \Gamma$ is \begin{equation}
\sum_{\vec \lambda} \chi_{\vec \lambda}(\vec x) \overline{\chi_{\vec \lambda}(\vec y)} 
=\sum_{\vec \lambda} \chi_{\vec \lambda}(\vec x) \chi_{\vec \lambda}(\vec{y}^{-1}) .
\end{equation} This is the properly-normalized delta-function for the Haar measure \eqref{Haar} of $\Gamma$, and
is  the 2d Yang-Mills propagator 

Now let us say $\Gamma$ is simply-laced, and  include the action of an outer-automorphism $\sigma$ on $\Gamma$. 
The outer-automorphism $\sigma$ can send $R_{\vec \lambda}$ to a different representation $R_{{\vec \lambda}'}$, in which case $R_{\vec \lambda}$ doesn't contribute to the trace anymore, or $R_{\vec \lambda}$ is sent to itself $R_{\vec \lambda}$, by an action $\cP_\sigma$ on $R_{\vec \lambda}$, in which case $R_{\vec \lambda}$ contributes to the trace by $\tr_{R_\sigma} \cP_\sigma x$. 
Then the trace is \begin{equation}
\sum_{{\vec \lambda}=\sigma({\vec \lambda})} \left(\tr_{R_{\vec\lambda}} \cP_\sigma \vec{x}\right)\left( \tr_{R_{\vec\lambda}} \cP_\sigma \vec{y}^{-1}\right).\label{twistedprop}
\end{equation} Now, in \cite{Fuchs:1995zr} it was proved that 
\begin{itemize}
\item the set ${\vec \lambda}=\sigma({\vec \lambda})$ of the irreps of $\Gamma$ is naturally identified with the set $\underline{{\vec \lambda}}$ of the irreps of $G$, and 

\item $\tr_{R_{\vec \lambda}} \cP_\sigma \vec{x}= \underline{\chi}_{\underline{{\vec \lambda}}}(\vec x)$ under this correspondence, where $\underline{\chi}_{\underline{{\vec \lambda}}}$ is the character of $G$ of the irrep $\underline{R}_{\underline{{\vec \lambda}}}$.
\end{itemize}
Then we conclude that \eqref{twistedprop} is equal to  \begin{equation}
\sum_{\underline{\vec \lambda}} \underline\chi_{\underline{\vec \lambda}}(\vec x)  \underline\chi_{\underline{\vec \lambda}}(\vec{y}^{-1}),
\end{equation} which is the propagator of the 2d Yang-Mills of non-simply-laced $G$. 

\section{Computations of Macdonald polynomials}\label{computation}
We want to explicitly find the Macdonald polynomials $P_{\vec {\vec \lambda}}$ that are orthogonal with respect to the Macdonald measure.
The Macdonald polynomial has a feature that
\begin{equation}
P_{\vec {\vec \lambda}}(\vec{a} ; q,t)=\chi_{\vec {\vec \lambda}}(\vec a) + \sum_{\vec \mu<\vec\lambda} f_{\vec \mu}(q,t) \chi_{\vec \mu}(\vec a),
\end{equation}
where the sum over ${\vec \mu}$ is only for ${\vec \mu}<{ {\vec \lambda}}$, in the sense that
\begin{equation}
{\vec {\vec \lambda}}- {\vec \mu} = \text{a sum of positive number of simple roots}.
\end{equation}

To achieve this, we just have to do the Schmidt orthogonalization with respect to the Macdonald measure, starting from
\begin{equation}
\chi_{{\vec {\vec \lambda}}_1}, \chi_{{\vec {\vec \lambda}}_2}, \ldots
\end{equation}
such that ${\vec {\vec \lambda}}_i < {\vec {\vec \lambda}}_j $ whenever $i<j$.

In practice, to obtain such an ordered set, one can compute the inner products $w_i = {\vec {\vec \lambda}}_i \cdot {\vec {\vec \rho}}$, where ${\vec {\vec \rho}}$ is the Weyl vector of $C_\ell$, and perform the sorting 
\begin{equation}
\chi_{w_1}, \chi_{w_2}, \ldots
\end{equation}
such that $w_i < w_j$ whenever $i<j$.

Also, for the Macdonald polynomials of type $(C_1^\vee,C_1)$ (which are also called the 1-variable Koornwinder polynomials, or the Askey-Wilson polynomials) we can use the explicit formula in terms of 
$q$-hypergeometric functions \cite{KoornwinderScholarpedia}. 

\section{The rule for the prefactor $\tilde K_{\Lambda}(a)$}\label{rule}
Consider a puncture of type $[l_1,l_2,\cdots,l_k]$ such that $\sum l_i=N$.
Denote the original mass parameters by $a_1,a_2,\ldots$ such that $a_1^{l_1} a_2^{l_2} \cdots a_k^{l_k}=1$. 
To twist by $\bZ_2$, we need to further impose the constraint  that \begin{equation}
(a_1t^{\frac{l_1-1}2}, a_1t^{\frac{l_1-3}2},\ldots,a_1t^{\frac{1-l_1}2},
a_2t^{\frac{l_2-1}2},\ldots,a_2t^{\frac{1-l_2}2},
\ldots,
a_kt^{\frac{l_k-1}2},\ldots,a_kt^{\frac{1-l_k}2}).
\end{equation}
is invariant under the $\bZ_2$ operation. We also need to include the factors of $\pm 1$ and $\pm i$ coming from the matrix $B_\sigma$ in \eqref{invariance}.
Then, the prefactor is given by multiplying the following factors: 
\begin{itemize}
\item An overall factor of $(-q;q)$.
\item For each $i<j$ such that $m=l_i \neq n=l_j$, \begin{equation}
\PE\left[
u^{0,1}\frac{t^{m-n+2}+t^{m-n+4}+\cdots+t^{m+n}}{1-q^2}\left(\frac{a_m^2}{a_n^2}+\frac{a_n^2}{a_m^2}\right)
\right]_{u=-1}
\end{equation}
Here $u^{0,1}$ stands for either $u$ or $1$.
\item Say the columns from $i$ to $j$ all have height $m$. 
\begin{itemize}
\item When the number of columns is even, $i-j+1=2\ell$. Call the mass parameters $z_1{}^{\pm1},\ldots,z_\ell{}^{\pm1}$.
\begin{multline}
\frac1{\prod_{n=1}^m ((-t)^n;q)^\ell(-(-t)^n;q)^\ell } 
\PE \bigg[ \\
+\frac{t^{2}+t^{4}+\cdots+t^{2m}}{1-q^2}E_2(\vec z) 
+u^{0,1}\frac{ut+(ut)^2+\cdots+(ut)^{m}}{1-q}E_1(\vec z)
\bigg]
\end{multline}
\item When the number of columns is odd, $i-j+1=2\ell+1$. Do include the case $\ell=0$. Call the mass parameters $z_1{}^{\pm1},\ldots,z_\ell{}^{\pm1},1$.
\begin{multline}
\frac1{\prod_{n=1}^m ((-t)^n;q)^{\ell+1}(-(-t)^n;q)^\ell } 
\PE \bigg[ \\
+\frac{t^{2}+t^{4}+\cdots+t^{2m}}{1-q^2} E_2(\vec z) 
+\frac{t^{2}+t^{4}+\cdots+t^{2m}}{1-q^2}E_1(\vec z) \\
+u^{0,1}\frac{ut+(ut)^2+\cdots+(ut)^{m}}{1-q}E_1(\vec z)
\bigg] \ . 
\end{multline}
\end{itemize}
\end{itemize}
It should be possible to determine the signs by carefully studying the action of $\cP_\sigma$ and $B_\sigma$ on the Slodowy slices. 

\section{Other free hypermultiplets}\label{other}
\newcommand{\nocontentsline}[3]{}
\let\addcontentsline\nocontentsline

\subsection{A sphere with punctures $[2,2]$, $[2,1,1]$ and $[1,1,1,1]$}
Without the twisting, in terms of chiral superfields,
we have matter in the following representations of $\SU(2)\times \SU(2) \times \SU(4)$: $[1; 0 ;1,0,0]$, $[1;0;0,0,1]$ and $[0;1;0,1,0]$, see \cite{Nanopoulos:2009xe,Chacaltana:2010ks}.

The only possible $\BZ_2$ action exchanges $\SU(4)$ representation $[1,0,0]$ with $[0,0,1]$.
Up to this point, we are still keeping all the fugacities of $\SU(4)$, namely $z_1, z_2, z_3$.
Note the characters of the following representations
\bea
\chi^{A_3}_{[1,0,0]} (\vec z)=\sum_{i=1}^3 z_i~,\quad  \chi^{A_3}_{[0,0,1]}(\vec z)=\sum_{i=1}^3 z_i^{-1}~, \quad \chi^{A_3}_{[0,1,0]}(\vec z)=\sum_{1\leq i<j \leq 3}z_i z_j~. 
\eea
The $\BZ_2$ operation thus sends $z_i \rightarrow 1/z_i$.
Next, we set
\bea
z_4=z_1^{-1}, \qquad z_3=z_2^{-1}~.  \label{replsu4}
\eea

The twisted character for the chiral fields in $[1;0;1,0,0]+[1;0;0,0,1]$ of $\SU(2)\times \SU(2) \times \SU(4)$ can be computed using the replacement \eref{replsu4} and taking into account of the doubling; we have
\bea
&\chi^{A_1}_{[1]}(\vec x)  \chi^{A_3}_{[1,0,0]}(\vec z) + \chi^{A_1}_{[1]}(\vec x)  \chi^{A_3}_{[0,0,1]}(\vec z) = (x+x^{-1}) \sum_{i=1}^3 (z_i+z_i^{-1}) \nn \\
&\rightarrow \quad (x^2+x^{-2}) \sum_{i=1}^3 (z_i^2+z_i^{-2}) = ( x^2+x^{-2}) E_1 (\vec z) ~.
\eea
Let us now consider the chiral fields in $[0;1;0,1,0]$ of $\SU(2)\times \SU(2) \times \SU(4)$. Under the replacement \eref{replsu4} (without any doubling at this stage), we have
\bea
\chi^{A_1}_{[1]}(\vec y)  \chi^{A_3}_{[0,1,0]} (\vec z)  \quad &\rightarrow \quad 2(y+y^{-1}) + ( y+y^{-1}) \sum_{\epsilon_i = \pm1} z_1^{\epsilon_1} z_2^{\epsilon_2} \nn \\
&= \quad 2(y+y^{-1}) + (y+y^{-1}) \widehat{E}_2(\vec z)~. \label{antisymtwist}
\eea
Since the representation $[0,1,0]$ of $\SU(4)$ is strictly real, there is no doubling for the second term $(y+y^{-1}) \widehat{E}_2(\vec z)$ of \eref{antisymtwist}.  However, there is still a doubling for the for the first term $2(y+y^{-1})$ of \eref{antisymtwist}; this leads to $(y^2+ y^{-2})$.

Thus, we have the twisted index
\bea
\tI_{[2,2],[2,1,1],[1^4]} &= \PE \Bigg[
\frac{ut}{1-q^2}(x^2+x^{-2})E_1(\vec z)  + \frac{t}{1-q^2}(y^2+y^{-2})   \nn \\
& \qquad + \frac{ut^{1/2}}{1-q} (y+y^{-1})\widehat{E}_2(\vec z)  
\Bigg]_{u=-1}~.
\eea
This can be written in the following form
\bea
\tI_{[2,2],[2,1,1],[1^4]} =   \frac{\tK_{[2,2]} \tK_{[2,1,1]} \tK_{\rm full}}{\tK_{\vec \rho}} 
\sum_{\vec \lambda} \frac{
\underline{\tP}_{\vec \lambda}(\vec x)
\underline{\tP}_{\vec \lambda}(\vec y)
\underline{\tP}_{\vec \lambda}(\vec z)}{\underline{\tP}_{\vec \lambda}(t^{\vec \rho})}~,
\eea
where $\tK_{\vec \Lambda}$, $\vec x$ and $\vec y$ are given below. 

\subsubsection{Puncture $[2,2]$}
For the puncture $[2,2]$,  the untwisted fugacity assignment is $(t^{1/2}x,t^{1/2}/x,t^{-1/2}x,t^{-1/2}/x)$. Therefore, the profile $\vec{\vec \Lambda}(\tau_+)$ has the form \begin{equation}
{\vec \Lambda}(\tau_+)=\begin{pmatrix}
0 & 0  & \bullet & 0 \\
0 & 0  & 0 &  \bullet\\
0 & 0  & 0 &0 \\
0 & 0  & 0 & 0.
\end{pmatrix} 
\end{equation} Comparing with \eqref{explicitZ2actionAodd} and Fig.~\ref{action}, we find \begin{equation}
B_\sigma=\diag(i,i,-i,-i).
\end{equation}
Then the argument of $\underline{\tP}$ after the twist is \begin{equation}
\vec x= (ix t^{1/2},it^{1/2}/x)
\end{equation} while
the prefactor is \begin{equation}
\tK_{[2,2]}(\vec x)=\frac{(-t;q)}{\{(-t;q)(t;q)\} \{(t^2;q)(-t^2;q)\}}\PE \left[ \left(\frac{t}{1-q} + \frac{ut^2}{1-q^2} \right) (x^2+x^{-2}) \right]_{u=-1}~.
\end{equation}

\subsubsection{Puncture $[2,1,1]$}
For the puncture $[2,1,1]$, 
the untwisted fugacity assignment is $(t^{1/2},y,1/y,t^{-1/2})$. Therefore, the profile $\vec{\vec \Lambda}(\tau_+)$ has the form \begin{equation}
{\vec \Lambda}(\tau_+)=\begin{pmatrix}
0 & 0  & 0 & \bullet \\
0 & 0  & 0 &  0\\
0 & 0  & 0 &0 \\
0 & 0  & 0 & 0
\end{pmatrix} .
\end{equation} Comparing with \eqref{explicitZ2actionAodd} and Fig.~\ref{action}, we find that $B_\sigma$ is trivial.
Then the argument of $\underline{\tP}$ is \begin{equation}
\vec y=(y, t^{1/2})
\end{equation} while
the prefactor is \begin{equation}
\tK_{[2,1,1]}(\vec y)=\frac{(-t;q)}{\{(-t;q)(t;q)\} \{(-t;q)(t^2;q)\}}\PE \left[ \left(\frac{t}{1-q} + \frac{t^3}{1-q^2} \right) (y^2+y^{-2}) \right]~.
\end{equation}

\subsection{A sphere with punctures $[3,2]$, $[2,2,1]$ and $[1,1,1,1,1]$}
Without the twisting, in terms of chiral superfields,
we have matter in the following representations of $\SU(2)\times \SU(5)$: $[1;1,0,0,0]$, $[1;0,0,0,1]$, $[0;0,1,0,0]$ and $[0;0,0,1,0]$.

The only possible $\BZ_2$ action exchanges  $[0,1,0,0]$ with $[0,0,1,0]$, and $[1,0,0,0]$ with $[0,0,0,1]$.
Up to this point, we are still keeping all the fugacities of $\SU(5)$, namely $z_1, \ldots, z_5$.
Note the characters of the following representations
\bea
\begin{array}{rclllcr}
\chi^{A_4}_{[1,0,0,0]}(\vec z)&=&\sum_{i=1}^5 z_i^1~,\qquad  &\chi^{A_4}_{[0,0,0,1]}(\vec z)&=&\sum_{i=1}^5 z_i^{-1}~,  \\
\chi^{A_4}_{[0,1,0,0]}(\vec z)&=&\sum_{1\leq i<j \leq 5}z_i z_j~, \qquad &\chi^{A_4}_{[0,0,1,0]}(\vec z)&=&\sum_{1\leq i<j \leq 5} z_i^{-1} z_j^{-1}~.
\end{array}
\eea
The $\BZ_2$ operation thus sends $z_i \rightarrow 1/z_i$.
Next, we set
\bea
z_5=z_1^{-1}, \qquad z_4=z_2^{-1}~, \qquad z_3=1~. \label{repl1}
\eea
Hence, under the replacement \eref{repl1} and taking into account of the doubling, we have
\bea
\chi^{A_1}_{[1]}(\vec y)  \chi^{A_5}_{[1,0,0,0,0]}(\vec z)+\chi^{A_1}_{[1]}(\vec y)  \chi^{A_5}_{[0,0,0,0,1]}(\vec z) \quad &\rightarrow \quad (y^2+y^{-2}) (E_1 (\vec z)+1) ~, \nn \\
\chi^{A_5}_{[0,1,0,0,0]} (\vec z)+\chi^{A_5}_{[0,0,0,1,0]} (\vec z)  \quad &\rightarrow \quad E_1(\vec z)+ E_2(\vec z)+1~. 
\eea

The twisted index is given by
\bea
\tI_{[3,2], [2,2,1],[1^5]} = \PE \left[ \frac{ut}{1-q} \left\{  (y^2+y^{-2})(E_1(\vec z)+1) + E_1(\vec z)+ E_2(\vec z)+1 \right \} \right]_{u=-1}~.
\eea
This can be written in the following form
\bea
\tI_{[3,2], [2,2,1],[1^5]} = \frac{\tK_{[3,2]} \tK_{[2,2,1]} \tK_{\rm full}}{\tK_{\vec \rho}} 
\sum_{\vec \lambda} \frac{
\underline{\tP}_{\vec \lambda}(\vec x)
\underline{\tP}_{\vec \lambda}(\vec y)
\underline{\tP}_{\vec \lambda}(\vec z)}{\underline{\tP}_{\vec \lambda}(t^{\vec \rho})}~,
\eea
where $\tK_{\vec \Lambda}$, $\vec x$ and $\vec y$ are given below. 

\subsubsection{Puncture $[3,2]$}
The argument of $\underline{\tP}$ is \begin{equation}
\vec x = (i t^{1/2}, t)~,
\end{equation} 
while the prefactor is 
\bea
\tK_{[3,2]} &= \frac{(-t;q)}{\{\prod_{i=1}^3 ((-t)^i;q) \}\{\prod_{j=1}^2 ((-t)^j;q) \}} \PE \left[ \frac{u(t^3+t^5)}{1-q^2} \right]_{u=-1}  \nn \\
&= \frac{(-t;q)}{\{\prod_{i=1}^3 ((-t)^i;q) \}\{\prod_{j=1}^2 ((-t)^j;q) \}\{(-t^3;q^2) (-t^5;q^2)\}}~.
\eea

\subsubsection{Puncture $[2,2,1]$}
The argument of $\underline{\tP}$ is \begin{equation}
\vec y = (i y t^{1/2}, i y^{-1} t^{1/2})~,
\end{equation} 
while the prefactor is 
\bea
\tK_{[2,2,1]} &= \frac{(-t;q)}{\{\prod_{i=1}^2 ((-t)^i;q)(-(-t)^i;q) \} \{ (-t;q) \}} \times \nn \\
& \qquad \PE \left[ \left\{\frac{u(t+t^2)}{1-q} +\frac{ut^3}{1-q^2} \right\} (y^2+y^{-2}) \right]_{u=-1}~.
\eea

\subsection{A sphere with punctures $[3,3]$, $[3,2,1]$ and $[1,1,1,1,1,1]$}
Without the twisting, in terms of chiral superfields,
we have matter in the following representations of $\SU(2)\times \SU(6)$: $[1;1,0,0,0,0]$, $[1;0,0,0,0,1]$, $[0;0,1,0,0,0]$ and $[0;0,0,0,1,0]$.

The only possible $\BZ_2$ action exchanges  $[0,1,0,0,0]$ with $[0,0,0,1,0]$, and $[1,0,0,0,0]$ with $[0,0,0,0,1]$.
Up to this point, we are still keeping all the fugacities of $\SU(6)$, namely $z_1, \ldots, z_6$.
Note the characters of the following representations
\bea
\begin{array}{rcrrcr}
\chi^{A_5}_{[1,0,0,0,0]}&=&\sum_{i=1}^6 z_i~,\qquad  \chi^{A_5}_{[0,0,0,0,1]}&=&\sum_{i=1}^6 z_i^{-1}~,  \\
\chi^{A_5}_{[0,1,0,0,0]}&=&\sum_{1\leq i<j \leq 6}z_i z_j~, \qquad
\chi^{A_5}_{[0,0,0,1,0]}&=&\sum_{1\leq i<j \leq 6} z_i^{-1} z_j^{-1}~.
\end{array}
\eea
The $\BZ_2$ operation thus sends $z_i \rightarrow 1/z_i$. Next, we set
\bea
z_6=z_1^{-1}, \qquad z_5=z_2^{-1}, \qquad z_4=z_3^{-1}~. \label{repl33}
\eea 

Hence, under the replacement \eref{repl33} and taking into account of the doubling, we have
\bea
\chi^{A_1}_{[1]}(\vec x)  \chi^{A_5}_{[1,0,0,0,0]}(\vec z)+\chi^{A_1}_{[1]}(\vec x)  \chi^{A_5}_{[0,0,0,0,1]}(\vec z) \quad &\rightarrow \quad (x^2+x^{-2}) E_1 (\vec z) ~, \nn \\
\chi^{A_5}_{[0,1,0,0,0]} (\vec z)+\chi^{A_5}_{[0,0,0,1,0]} (\vec z)  \quad &\rightarrow \quad E_2(\vec z)+3~. 
\eea
The twisted index is
\bea
\tI_{[3,3],[3,2,1],[1^6]} &= \PE \left[ \frac{ut}{1-q} \left\{ (x^2 +x^{-2}) E_1 (\vec z) +  (E_2(\vec z)+3) \right \} \right]_{u=-1}~.
\eea
This can be written in the following form
\bea
\tI_{[3,3],[3,2,1],[1^6]} = \frac{\tK_{[3,3]} \tK_{[3,2,1]} \tK_{\rm full}}{\tK_{\vec \rho}} 
\sum_{\vec \lambda} \frac{
\underline{\tP}_{\vec \lambda}(\vec x)
\underline{\tP}_{\vec \lambda}(\vec y)
\underline{\tP}_{\vec \lambda}(\vec z)}{\underline{\tP}_{\vec \lambda}(t^{\vec \rho})}~,
\eea
where $\tK_{\vec \Lambda}$, $\vec x$ and $\vec y$ are given below.

\subsubsection{Puncture $[3,3]$}
The argument of $\underline{\tP}$ is \begin{equation}
\vec x= (x t,~x^{-1} t,~ x)~,
\end{equation} 
while the prefactor is \begin{equation}
\tK_{[3,3]}= \frac{(-t;q)}{\prod_{i=1}^3 ((-t)^i;q) (-(-t)^i;q)} \PE \left[(x^2 + x^{-2}) \left \{  \frac{t}{1-q} + \frac{ut^2}{1-q} + \frac{t^3}{1-q} \right\} \right]_{u=-1}~.
\end{equation}

\subsubsection{Puncture $[3,2,1]$}
The argument of $\underline{\tP}$ is \begin{equation}
\vec y=(it, ~ t^{1/2} ,~ i)~,
\end{equation} 
while the prefactor is 
\bea
\tK_{[3,2,1]}&= \frac{(-t;q)}{\left \{ \prod_{i=1}^3 ((-t)^i;q) \right\} \{ \prod_{j=1}^2 ((-t)^j;q) \} \{ (-t; q) \} }  \times \nn \\
& \qquad \PE \left[ \frac{u(t^3+t^5)}{1-q^2} \mathbf{+} \frac{t^4}{1-q^2} + \frac{ut^3}{1-q^2}  \right ]_{u=-1} \nn \\
&= \frac{(-t;q)}{\left \{ \prod_{i=1}^3 ((-t)^i;q) \right\} \{ \prod_{j=1}^2 ((-t)^j;q) \} \{ (-t; q) \} \{ (-t^3; q^2)^2  (t^4;q^2) (-t^5; q^2) \}}~.
\eea

\subsection{A sphere with punctures $[4,2]$, $[2,2,2]$ and $[1,1,1,1,1,1]$}
Without the twisting, in terms of chiral superfields,
we have matter in the following representations of $\SU(3) \times \SU(6)$: $[1,0;1,0,0,0,0]$, $[0,1;0,0,0,0,1]$, $[0,0;0,0,1,0,0]$, see \cite{Tachikawa:2011yr}.

The only possible $\BZ_2$ action exchanges $\SU(6)$ representation $[1,0,0,0,0]$ with $[0,0,0,0,1]$, and $\SU(3)$ representation $[1,0]$ 
with $[0,1]$.
As before, the $\BZ_2$ operation thus sends $z_i \rightarrow 1/z_i$ (where $z_1, \ldots, z_6$ are $\SU(6)$ fugacities) 
and $y_i \rightarrow 1/y_i$ (where $y_1,y_2, y_3$ are $\SU(3)$ fugacities). 
Next, we set
\bea
z_4=z_1^{-1}, \qquad z_5=z_2^{-1}, \qquad z_6=z_3^{-1}~. \label{replsu6}
\eea
Similarly for $\SU(3)$, we set
\bea
y_1= y, \qquad y_2 = 1, \qquad y_3^{-1} = y~. \label{replsu3}
\eea

The twisted character for the chiral fields in $[1,0;1,0,0,0,0]+[0,1;0,0,0,0,1]$ of $\SU(3) \times \SU(6)$ can be computed using the replacements \eref{replsu6}, \eref{replsu3} and taking into account of the doubling; we have
\bea
\chi^{A_2}_{[1,0]}(\vec y)  \chi^{A_5}_{[1,0,0,0,0]}(\vec z)+\chi^{A_2}_{[0,1]}(\vec y)  \chi^{A_5}_{[0,0,0,0,1]}(\vec z) \quad &\rightarrow \quad (y^2+1+y^{-2}) E_1 (\vec z) ~.
\eea
Let us now consider the chiral fields in $[0,0,1,0,0]$ of $\SU(6)$. Under the replacement \eref{replsu6} (without any doubling at this stage), we have
\bea
\chi^{A_5}_{[0,0,1,0,0]} (\vec z)  \quad &\rightarrow \quad 2\sum_{i=1}^3 (z_i+z_i^{-1}) +\sum_{\epsilon_i = \pm1} z_1^{\epsilon_1} z_2^{\epsilon_2} z_3^{\epsilon_3} = 2E_1(z_{1/2}) +E_3(\vec z)  \quad ~. \label{antisymtwistA5}
\eea
Since the representation $[0,0,1,0,0]$ is real, there is no doubling for the second term $E_3(\vec z)$ of \eref{antisymtwistA5}. However, there is still a doubling for the for the first term $2E_1(z_{1/2})$ of \eref{antisymtwistA5}; this leads to $E_1(\vec z)$.

Thus, the twisted index is
\bea
\tI_{[4,2],[2,2,2],[1^6]} &=\PE \Big[  \frac{t}{1-q^2} (y^2+y^{-2}+1) E_1(\vec z)+ \frac{t}{1-q^2} E_1(\vec z)    \nn \\
& \quad \qquad +\frac{ut^{1/2}}{1-q} E_3(\vec z) \Big]_{u=-1}~.
\eea
This can be written in the following form
\bea
\tI_{[4,2],[2,2,2],[1^6]} = \frac{\tK_{[4,2]} \tK_{[2,2,2]} \tK_{\rm full}}{\tK_{\vec \rho}} 
\sum_{\vec \lambda} \frac{
\underline{\tP}_{\vec \lambda}(\vec x)
\underline{\tP}_{\vec \lambda}(\vec y)
\underline{\tP}_{\vec \lambda}(\vec z)}{\underline{\tP}_{\vec \lambda}(t^{\vec \rho})}~,
\eea
where $\tK_{\vec \Lambda}$, $\vec x$ and $\vec y$ are given below.

\subsubsection{Puncture $[4,2]$}
The argument of $\underline{\tP}$ is \begin{equation}
\vec x=(t^{3/2},t^{1/2},t^{1/2})
\end{equation} while the prefactor is \begin{equation}
\tK_{[4,2]}=(-t;q) \frac{1}{(t^4;q^2)(t^6;q^2)}\prod_{i=1}^4 \frac{1}{((-t)^i;q)} \prod_{i=1}^2 \frac{1}{((-t)^i;q)}~.
\end{equation}

\subsubsection{Puncture $[2,2,2]$}
The argument of $\underline{\tP}$ is \begin{equation}
\vec y=(yt^{1/2},y^{-1}t^{1/2},-t^{1/2})
\end{equation} 
while the prefactor is 
\bea
\tK_{[2,2,2]}(\vec y) &= \PE \Big[\left(\frac{ut}{1-q}+ \frac{t^2}{1-q} + \frac{t^2}{1-q^2}+ \frac{t^2}{1-q^4} \right) E_1(\vec y)  \Big]_{u=-1} \nn \\
& \qquad \times \frac{1}{(-t;q)^2(t^2;q)^2(t;q)(-t^2;q)}~.
\eea

\bibliographystyle{ytphys}
\small\baselineskip=.93\baselineskip
\bibliography{ref}

\end{document}